\documentclass[twocolumn,10pt]{siggraph}	
\usepackage{natbib}
\usepackage{amscd}
\usepackage{amsmath}
\usepackage{amsthm}
\usepackage{amsfonts}
\usepackage{amssymb}
\usepackage{graphicx}
\usepackage{url}
\usepackage{hyperref}

\newcommand{\0}{{\bm 0}}

\newcommand{\A}{\bm{A}}

\newcommand{\B}{\bm{B}}

\newcommand{\Bmatr}[1]{\begin{bmatrix}\displaystyle #1\end{bmatrix}}

\newcommand{\Cov}{\mbox{\rm Cov}}
\newcommand{\C}{\bm{C}}

\newcommand{\D}{\bm{D}}

\newcommand{\Fc}{{\cal F}}

\newcommand{\Gradient}{\nabla}

\newcommand{\I}{\bm{I}}

\newcommand{\K}{\bm{K}}
\newcommand{\Lam}{\bms{\Lambda}}

\newcommand{\Lm}{\bm{L}}

\newcommand{\M}{\bm{M}}
\newcommand{\Mmb}{\bar{\M}}

\newcommand{\Nc}{{\cal N}}

\newcommand{\Phim}{\bms{\Phi}}

\renewcommand{\P}{\bm{P}}
\newcommand{\Pm}{\bm{P}}

\newcommand{\Q}{\bm{Q}}

\newcommand{\R}{\bm{R}}
\renewcommand{\S}{\bm{S}}

\newcommand{\Sm}{\bm{S}}

\newcommand{\Var}{\mbox{\rm Var}}

\newcommand{\X}{\bm{X}}

\newcommand{\an}[1]{\begin{align}#1\end{align}}
\newcommand{\ab}[1]{\begin{align*}#1\end{align*}}

\newcommand{\av}{\bm{a}}
\newcommand{\desclist}[1]{\begin{description}#1\end{description}}

\newcommand{\itemlist}[1]{\begin{itemize}#1\end{itemize}}

\newcommand{\bms}[1]{{\boldsymbol{#1}}}
\newcommand{\bm}[1]{{\bf #1}}

\renewcommand{\c}{\bm{c}}
\newcommand{\ch}{\hat{c}}

\newcommand{\df}[2]{\displaystyle \frac{\mbox{\rm d} #1}{\mbox{\rm d} #2}}
\newcommand{\dfl}[2]{\mbox{\rm d} #1 / \mbox{\rm d} #2}

\newcommand{\diag}[1]{\mbox{ \bf diag}\matrx{#1}}

\newcommand{\disp}{{\rm D}}

\newcommand{\dspfrac}[2]{\frac{\displaystyle #1}{\displaystyle #2} }

\newcommand{\enumlist}[1]{\begin{enumerate}#1\end{enumerate}}

\newcommand{\epsi}{\bms{\epsilon}}
\newcommand{\eqdef}{:=}

\newcommand{\ev}{\bm{e}}
\newcommand{\evt}{{\,\ev^\tp}}
\newcommand{\e}{\bm{e}}

\newcommand{\goesto}{\rightarrow}

\newcommand{\hide}[1]{}	

\newcommand{\ie}{i.e.\ }

\newcommand{\la}{\!\leftarrow\!}

\newcommand{\matrx}[1]{{\left[ \stackrel{}{#1}\right]}}

\newcommand{\ov}{\overline}

\newcommand{\pf}[2]{\displaystyle \frac{\partial #1}{\partial #2}}

\newcommand{\pfinline}[2]{\partial #1 / \partial #2}

\newcommand{\piv}{\bms{\pi}}

\newcommand{\pr}{\protect}

\newcommand{\p}{\bm{p}}

\newcommand{\rec}{{\rm R}}

\newcommand{\sel}{{\rm S}}

\newcommand{\suchthat}{\colon}

\newcommand{\tb}{\!\! & \!\!}

\newcommand{\tp}{\top}	

\newcommand{\tr}{\!^\top}

\newcommand{\uv}{\bm{u}}
\newcommand{\uvt}{\uv^\tp}

\newcommand{\vv}{\bm{v}}

\newcommand{\wb}{\ov{w}}

\newcommand{\xiv}{\bms{\xi}}

\newcommand{\xvh}{\hat{\x}}

\newcommand{\x}{\bm{x}}

\newcommand{\y}{\bm{y}}
\newcommand{\zetav}{\bms{\zeta}}

\newcommand{\z}{\bm{z}}
\newfont{\gilfont}{cmsy10 scaled\magstep0}
\newcommand{\Reals}{\mathbb{R}} 

\newtheorem{Theorem}{Theorem}

\newtheorem*{Theorem*}{Theorem}
\newtheorem*{Conjecture*}{Conjecture}
\newtheorem*{Corollary*}{Corollary}
\newtheorem*{Proposition*}{Proposition}
\newtheorem*{Definition*}{Definition}
\newtheorem*{Lemma*}{Lemma}
\newtheorem*{Paradox*}{Paradox}
\newtheorem*{Principle*}{Principle}
\newtheorem*{Recursion*}{Recursion}
\newtheorem*{Result*}{Result}

\newtheorem{Theorem:}[Theorem]{Theorem:}
\newtheorem{Conjecture:}[Theorem]{Conjecture:}
\newtheorem{Corollary:}[Theorem]{Corollary:}
\newtheorem{Definition:}[Theorem]{Definition:}
\newtheorem{Lemma:}[Theorem]{Lemma:}
\newtheorem{Paradox:}[Theorem]{Paradox:}
\newtheorem{Principle:}[Theorem]{Principle:}
\newtheorem{Proposition:}[Theorem]{Proposition:}
\newtheorem{Recursion:}[Theorem]{Recursion:}
\newtheorem{Result:}[Theorem]{Result:}

\newtheorem{Theorem-InOrder}[Theorem]{Theorem}
\newtheorem{Conjecture-InOrder}[Theorem]{Conjecture}
\newtheorem{Corollary-InOrder}[Theorem]{Corollary}
\newtheorem{Definition-InOrder}[Theorem]{Definition}
\newtheorem{Lemma-InOrder}[Theorem]{Lemma}
\newtheorem{Paradox-InOrder}[Theorem]{Paradox}
\newtheorem{Principle-InOrder}[Theorem]{Principle}
\newtheorem{Proposition-InOrder}[Theorem]{Proposition}
\newtheorem{Recursion-InOrder}[Theorem]{Recursion}
\newtheorem{Result-InOrder}[Theorem]{Result}
\newtheorem{Counterexample-InOrder} [Theorem]{Counterexample}

\theoremstyle{plain}

\theoremstyle{plain}

\newtheorem*{OpenQuestion}{Open Question}
\newcommand{\McD}{McNamara:and:Dall:2011}
\newcommand{\McDT}{\citet{McNamara:and:Dall:2011}}

\newcommand{\McDA}{\citeauthor{\McD}}
\newcommand{\FAC}{{\cal F\!A}}
\newcommand{\fac}{fitness-abundance covariance}

\begin{document}
\title{The Evolution of Dispersal in Random Environments and \\
The Principle of Partial Control
}
\author{Lee Altenberg \\ \url{altenber@hawaii.edu}}

\maketitle

\begin{abstract}
\McDT\ identified novel relationships between the abundance of a species in different environments, the temporal properties of environmental change, and selection for or against dispersal.  Here, the mathematics underlying these relationships in their two-environment model are investigated for arbitrary numbers of environments.  The effect they described is quantified as the \fac.  The phase in the life cycle where the population is censused is crucial for the implications of the  \fac.  These relationships are shown to connect to the population genetics literature on the Reduction Principle for the evolution of genetic systems and migration.  Conditions that produce selection for increased unconditional dispersal are found to be new instances of departures from reduction described by the ``Principle of Partial Control'' proposed for the evolution of modifier genes.  According to this principle, variation that only partially controls the processes that transform the transmitted information of organisms may be selected to increase these processes.  Mathematical methods of Karlin, Friedland, and Elsner, Johnson, and Neumann, are central in generalizing the analysis.\footnote{Dedicated to the memory of Professor Michael Neumann, one of whose many elegant theorems provides for a result presented here.}  Analysis of the adaptive landscape of the model shows that the evolution of conditional dispersal is very sensitive to the spectrum of genetic variation the population is capable of producing, and suggests that empirical study of particular species will require an evaluation of its variational properties.
\end{abstract}
\sloppy
\section{Introduction}

In analyzing a model of a population that disperses in a patchy environment subject to random environmental change, \citet{\McD} describe ``how an underappreciated evolutionary process, which we term `The Multiplier Effect', can limit the evolutionary value of responding adaptively to environmental cues, and thus favour the evolutionary persistence of otherwise paradoxical unconditional strategies.''  By ``multiplier effect'', \McDA\ mean, 
\begin{quote}
If a genotype is distributed in space and its fitness varies with
location, then selection will change the spatial distribution of the
genotype through its effect on population demography. This process
can accumulate genotype members in locations to which they are well
suited. This accumulation by selection is the multiplier effect.
\end{quote}
It is possible, they discover, for the `multiplier effect' to reverse --- for there to be an excess of the population in the \emph{worst} habitats --- when there is very rapid environmental change.  The environmental change they model is a Markov process where are large number of patches switch independently between two environments that produce different growth rates for a population of organisms.  They find that for moderate rates of environmental change, populations will have higher asymptotic growth rates if they reduce their rate of unconditional dispersal between patches, which produces effective selection for lower dispersal.

Their key finding is that the reversal of the `multiplier effect' due to rapid environmental change corresponds exactly with a reversal in the direction in which dispersal evolves:  when abundance is greater on better habitats, lower dispersal is selected for; when abundance is greater on worse habitats because the environment changes so fast, there is selection for higher dispersal.

\McDA\ conclude their paper saying, ``the multiplier effect may underpin the evolution and maintenance of unconditional strategies in many biological systems.''  This is indeed the case.  Their results are in fact part of the phenomenon already known as the ``reduction principle'', which was first described as such in models for the evolution of linkage \citep{Feldman:1972}, and subsequently extended to models for the evolution of mutation rates, gene conversion, dispersal, sexual reproduction, and even cultural transmission of traditionalism \citep{Altenberg:1984}.  The reduction principle also underlies other phenomena:  the `error catastrophe' in quasispecies dynamics, and the effect of population subdivision on the maintenance of genetic diversity.  

The Reduction Principle can be stated, in a rather general form, as the widely exhibited phenomenon that \emph{mixing reduces growth}, and \emph{differential growth selects for reduced mixing}.  

While the reduction phenomenon studied in \citet{\McD} is not a new concept, three particular aspects of their study are novel:  
\enumlist{
\item their discovery of conditions that cause mixing to \emph{increase} growth --- which addresses the open problem posed in \citet[Open Question 3.1]{Altenberg:2004:Open} as to the conditions that produce departures from the reduction principle;
\item that these departures from reduction emerge from very rapidly changing environments; and
\item that these departures from reduction correspond to reversals in the association between fitness and abundance in different environments.  
}

\McDA\ produce these results from a two-environment model.  A principal goal here is to generalize each of these findings to arbitrary numbers of environments.  Insight on how to generalize them is provided by clues in their results.  Some of these clues point to the main tool used to achieve the generalization, a theorem of the late Sam Karlin, to be described.  

The property described by \McDA\ as `the multiplier effect' is here made mathematically precise, as a positive covariance between \emph{fitness} and the \emph{excess of the stationary distribution} of the population above what it would be in the absence of differential growth rates, as censused just after dispersal.  I refer to this quantity as the \emph{fitness-abundance covariance}, which is a bit more descriptive and specific than the term `multiplier effect', which already has long use as a concept in economics.

A critical aspect to use of the fitness-abundance covariance is the phase in the life cycle at which the census is taken.  When \McDA\ say that ``individuals are likely to find themselves in circumstances to which
they are well-adapted,'' it matters where in its life cycle the individual finds itself  --- whether it is on its natal site or has already dispersed.  \McDA\ do not explicitly address the phase at which they take their census, but their model shows it to be just after dispersal, before reproduction. 

The issue of census phase is explicitly addressed here, and is shown to critically affect properties of the fitness-abundance covariance.  For populations censused just {after} dispersal, one cannot say in general that ``individuals are already likely to be on the better site.''  As a consequence of this phase dependence, a novel result found here is that by taking a census of the populations before and after reproduction, one can in certain situations infer a bound on the duration of changing environments.  

A result in \citet{\McD} that garnered considerable attention is that `` `stupid strategies' could be best for the genes''  \citep{ScienceDaily:2011-3-1:Stupid}:
\begin{quote}
One underappreciated consequence of the multiplier effect is that
because individuals tend to be in locations to which they are well
suited, its mere existence informs an organism that it is liable to be in
favourable circumstances. This information can outweigh environmental
cues to the contrary, so that an individual should place more
weight on the fact it exists than on any additional cues of location
quality.  \McDT
\end{quote}
The general analysis provided here produces results that seems to contradict the above:   philopatry is never an evolutionarily stable strategy when there is any level of environmental change;  it can always be invaded by organisms that disperse from the correct environments.  

In an attempt to resolve the apparent contradiction, I take a closer examination of the adaptive landscape --- the gradient of fitness over the space of conditional dispersal probabilities.  What is found is that the evolutionarily stable state is highly sensitive to constraints on the organismal variability for dispersal probabilities.  Slight changes in the constraints can shift the evolutionarily stable state from complete philopatry to complete dispersal from some environments.  This sensitivity means that conditional dispersal may be a highly volatile trait evolutionarily.  Moreover, to understand the evolution of any particular species requires an analysis of the constraints on the phenotype, and the probabilities of generating heritable variation in any phenotypic direction --- in short, an {evolvability analysis} \citep{Wagner:and:Altenberg:1996}.

While it is relatively straightforward to determine the long-term growth rates of different dispersal phenotypes, determining the likelihood that such phenotypes will be produced by the population plunges one into issues of the organism's perceptual and cognitive limits, ecological correlates, and the genotype-phenotype map, and requires specific empirical knowledge of the organism and its variability in order to address.  This is perhaps why, as  \citet[p. 494]{Levinton:1988} insightfully writes, ``Evolutionary biologists have been mainly concerned with the {\em fate of variability} in populations, not the {\em generation of variability}. \ldots Whatever the reason, the time has come to reemphasize the study of the origin of variation.''  A principle finding here is that the evolutionary outcome is not determined by the adaptive landscape studied here, and we are pointed instead to examine the variational properties of each particular species in question.

\subsection{The Reduction Principle and Fisher's Fundamental Theorem of Natural Selection}

The intuition as to why there should be selection for lower dispersal in a population at a stationary balance between dispersal and selection is well expressed in the following explanation:
\begin{quote}
Even in the absence of genetic variability for local adaptation in a spatially heterogeneous environment, migration will be selected against because on the average an individual will disperse to an environment worse than the one it was born in, since better environments harbor more individuals. \citep{Olivieri:Michalakis:and:Gouyon:1995}.
\end{quote}
This is a description of populations that have equilibrated to a balance between dispersal and differential growth.  Fisher's Fundamental Theorem is that differential growth rates increase the mean fitness of the population by an amount equal to the variance in the growth rates.  When the population is at a stationary distribution, however, this requires that dispersal \emph{decrease} the mean fitness by exactly the same amount.  Fisher uses the phrase ``deterioration of the environment'' (\citealt{Fisher:1958}; discussed in \citealt{Price:1972}) to describe this exact counterbalance to the variance in fitness that increases the mean fitness.  But he includes mutation in this concept:
\begin{quote}
\ldots an equilibrium must be established in which the
rate of elimination is equal to the rate of mutation.
To put the matter in another way we may say that each mutation of this kind is allowed to contribute exactly as much to the genetic variance of fitness in the species as will provide a rate of improvement equivalent to the rate of deterioration caused by the continual occurrence of the mutation. \citep[p. 41]{Fisher:1958}
\end{quote}
Fisher was thinking of mutation, not dispersal, in the above.  But as we shall see later, the same mathematics underlies both.  
Like a the mutation/selection balance Fisher describes, dispersal will generally be to lesser quality environments when the population has reached a growth/dispersal balance.  

The interchangeability of many results in population genetics between mutation and dispersal reflects the fact that an organism's location, like its genotype, is transmissible information about its state, and its degree of preservation  during  transmission is itself an organismal phenotype and subject to evolution (\citealt{Cavalli-Sforza:and:Feldman:1973:MCI,Karlin:and:McGregor:1974}; \citealt[pp. 15--16,  p. 178]{Altenberg:1984} \citealt{Schauber:Goodwin:Jones:and:Ostfeld:2007,Odling-Smee:2007}).   The issue of the faithfulness of transmission brings us to the reduction principle.

\section{A Review of the Reduction Principle}
\McDA\ are more correct than perhaps even they realized in noting that their subject is an ``underappreciated evolutionary process''.  It is clear that awareness of the body of population genetics literature on the reduction principle has not fully percolated between disciplines.   
Karlin's \citeyearpar{Karlin:1982} key theorem on the reduction phenomenon, and its application to the evolution of dispersal \citet{Altenberg:1984}, were independently duplicated recently by \citet{Kirkland:Li:and:Schreiber:2006}.  And \citet{\McD} were evidently unaware of the paper by \citet{Kirkland:Li:and:Schreiber:2006}, published in a mathematics journal.  

One main goal of this paper, therefore, is to provide a `portal' to the reduction principle, its historical development, and methods of analysis for a broader audience.  Here, I tie-in the work of \citet{\McD} to the larger stream of work on the reduction principle, and show that their work contributes toward answering one of the main open problems in the field:  how departures from the reduction phenomenon are produced.  

It may be appropriate to apologize for the density of equations in this paper, as equations nowadays are often being relegated to online-only supplements.  But the subject of this paper is in fact mathematical methodology.  It is the mathematics that creates a single conceptual and analytical framework for dispersal, recombination, mutation, random environments, and multiple genetic processes.  To show how they all share in a single body of results requires we delve into the mathematics.

It should be noted that many theoretical studies constrain their analysis to models having only $n=2$ patches or genotypes, to allow explicit calculation of the eigenvalues and eigenvectors (e.g. \citealt{\McD}, \citealt{Steinmeyer:and:Wilke:2009}).  There are mathematical tools from the reduction principle literature, however --- in particular the aforementioned theorems of Karlin --- that make analytical results tractable for arbitrary $n$.  Dissemination of these tools to a larger audience is another principal goal of this paper.  They are laid out in \emph{Methods}.

\subsection{Development of the Reduction Principle}

In the first analyses of genetic modifiers of mutation, recombination, and migration by Marc Feldman and coworkers in the 1970s, a common result kept appearing, which was that reduced levels of mutation, recombination, or migration would evolve when populations were near equilibrium under a balance between the forces of selection and transmission.  The earliest appearance of the reduction phenomenon in the literature is perhaps Fisher's \citeyearpar[p. 130]{Fisher:1930} assertion that ``the presence of pairs of factors in the same chromosome, the selective advantage of each of which reverses that of the other, will always tend to diminish recombination, and therefore to increase the intensity of linkage in the chromosomes of that species.''  This claim was mathematically verified by \citet{Kimura:1956}.  Nei \citeyearpar{Nei:1967:Modification,Nei:1969:Linkage} posed the first three-locus model for the evolution of recombination, with a modifier locus controlling the recombination between two loci under selection, and found that only reduced recombination would evolve.  The first fixed-point stability analysis of modifiers of recombination between two loci under viability selection was by \citet{Feldman:1972}, who found that recombination would be reduced by evolution.  Subsequent studies extended the reduction result to larger and larger spaces of models, including modifiers of:  
\desclist{
\item [{\it dispersal:}] \citet{Karlin:and:McGregor:1972:Modifier,Balkau:and:Feldman:1973,Karlin:and:McGregor:1974,Teague:1977,Asmussen:1983:Evolution,Hastings:1983,Feldman:and:Liberman:1986,Liberman:and:Feldman:1989,Wiener:and:Feldman:1991:Evolution,Wiener:and:Feldman:1993:Effects};
\item [{\it recombination:}]  \citet{Feldman:1972,Karlin:and:McGregor:1972:Modifier,Feldman:and:Balkau:1972,Feldman:and:Balkau:1973,Feldman:and:Krakauer:1976,Feldman:Christiansen:and:Brooks:1980,Liberman:and:Feldman:1986:Recombination,Feldman:and:Liberman:1986}; and
\item [{\it mutation:}]  \citet{Karlin:and:McGregor:1972:Modifier,Liberman:and:Feldman:1986:Mutation,Feldman:and:Liberman:1986}.
}
(Note that this literature prefers the term `migration', while `dispersal' is preferred in the ecology literature.  Literature searches need to include both.)

These studies also extended the generality of the reduction results to include arbitrary large modified rates, arbitrary viability selection regimes, and multiple modifier alleles.  They could only analyze the case of two patches or two alleles per selected locus, however, due to their use of closed-form solutions for the determinants or eigenvalues.  \citet{Hastings:1983} is notable in extending the phenomenon to continuous spatial variation.

\citet{Feldman:1972} proposed that the essential direction of evolution for the recombination modifiers was reduction in the recombination rates.  Shortly thereafter, \citet{Karlin:and:McGregor:1972:Modifier,Karlin:and:McGregor:1974} proposed an alternative idea, that the underlying governor for the direction of modifier evolution was the ``Mean Fitness Principle''.  The Mean Fitness Principle proposed that a modifier allele increases when rare if and only if it changes the parameter it controls to a value that would increase the mean fitness of the population at equilibrium.  Both reduction and mean fitness principles explained the known results at that time.  However, \citet[Fig. 1]{Karlin:and:Carmelli:1975} found an example where reducing recombination would \emph{decrease} the mean fitness of the population, while \citet{Feldman:Christiansen:and:Brooks:1980} showed that, even for this example, an allele reducing recombination would grow in the population.  Therefore, only the reduction principle remained unfalsified.  Subsequent modifier gene studies have found other counterexamples to the mean fitness principle \citep{Uyenoyama:and:Waller:1991:Coevolution-I,Uyenoyama:and:Waller:1991:Coevolution-II,Wiener:and:Feldman:1993:Effects}.  In \citet{Feldman:Christiansen:and:Brooks:1980} is where reduction was first referred to as a ``principle''.  

\subsection{Karlin's Theorems}

During the time period of these developments, Karlin had, ironically, elucidated the mathematical foundations for the reduction principle himself --- without realizing it.

Karlin was investigating a seemingly distant topic --- how population subdivision would affect the maintenance of genetic variation.  
To understand how the protection of alleles against extinction depended on migration patterns and rates, 
\citet{Karlin:1976,Karlin:1982} developed two general theorems on the spectral radius of perturbations of migration-selection systems.  The spectral radius is the growth rate for the whole group of genotypes that comprise the perturbation as they approach a stationary distribution among themselves.  

These theorems show how, for two different kinds of variation in migration, a greater level of `mixing' reduces the spectral radius of the stability matrix for the system, and thus may cause some alleles to lose their protection against extinction.  Hence, greater levels of mixing would lead to fewer polymorphic alleles.  Preparatory to this work was the paper by \citet{Friedland:and:Karlin:1975}.  The theorems first appear, without proof, in \citet[pp. 642--647]{Karlin:1976}, and with proof as Theorems 5.1 and 5.2 in \citet{Karlin:1982}, restated as follows:
\begin{Theorem-InOrder}[{\citealt[Theorem 5.1, pp. 114--116, 197--198]{Karlin:1982}}]  
\label{Theorem5.1} 
Consider a family of stochastic matrices that commute and are symmetrizable to positive definite matrices:
\an{
\label{eq:KarlinFamily}
\Fc \eqdef \{ \M_h = \Lm \Sm_h \R \colon \M_h \M_k = \M_k \M_h \},
}
where $\Lm$ and $\R$ are positive diagonal matrices, and each $\Sm_h$ is a positive definite symmetric real matrix.   Let $\D$ be a positive diagonal matrix.  Then
for each $\M_h, \M_k \in \Fc$, the spectral radius, $\rho$, satisfies:
\ab{
\rho(\M_h \M_k \D) \leq \rho(\M_k \D).
}
\end{Theorem-InOrder}

\begin{Theorem-InOrder}[{\citealt[Theorem 5.2, pp. 117--118, 194--196]{Karlin:1982}}]
Let $\M$ be a non-negative irreducible stochastic matrix.  Consider the family of matrices
\[
\M{(\alpha)} = (1-\alpha) \I + \alpha \M.
\]
Then for any positive diagonal matrix $\D$, the spectral radius
\[
\rho(\alpha) = \rho( \M{(\alpha)} \D )
\]
is decreasing as $\alpha$ increases (strictly provided $\D \neq d \I$).
\end{Theorem-InOrder}
In Theorem 5.1, `more mixing' is produced the application of a second mixing operator;  in Theorem 5.2, more mixing is produced by the equal scalar multiplication of all the transition probabilities between states.  In both cases, greater mixing reduces the spectral radius, which represents the asymptotic growth rate of a rare allele in  Karlin's analysis.

Theorems 5.1 and 5.2 display certain tradeoffs in generality.  In Theorem 5.2, $\M$ may be any irreducible stochastic matrix, but the variation in the matrix family is restricted to a single parameter --- the scaling of the transition probabilities.  In Theorem 5.1 on the other hand, the variation in the matrix family is more general in that it has up to $n-1$ degrees of freedom to vary (see \emph{Remark} for Lemma \ref{Lemma:CanonicalForm}), but the matrix class itself is narrower with the constraint that they be symmetrizable.

Karlin's proof of Theorem 5.2 relied upon the recently minted variational formula for the spectral radius of \citet{Donsker:and:Varadhan:1975}.  These results on the reduction principle, and their means of generalization, all came into being in the same time period.

\section{Application of Karlin's theorems to the Evolution of Dispersal and Genetic Systems}

My own contribution to the reduction principle began with a conjecture by Marcus Feldman (1980, personal communication).  The existence of polymorphisms for genes controlling recombination and mutation rates had been discovered theoretically by \citet{Feldman:and:Balkau:1973} and \citet{Feldman:and:Krakauer:1976}).  Generalizing from these examples, Feldman conjectured that whenever a parameter controlled by a gene enters \emph{linearly} into the recursion on the frequency dynamics, then a polymorphism for that gene would exist in which: 
\enumlist{
\item the population, when fixed on an allele producing a particular value of the linear parameter, is at an equilibrium;
\item each allele's average value of the parameter is equal to that particular value; and\label{item:VA}
\item the gene is in linkage equilibrium with the rest of the genome.
}
Because condition \ref{item:VA}.\ was analogous to the condition for alleles under viability selection that their marginal fitnesses be equal at equilibrium, these polymorphisms were called `viability-analogous, Hardy-Weinberg' (VAHW) modifier polymorphisms.

The repeated appearance of the VAHW polymorphisms, and the repeated occurrence of the reduction principle in models of different phenomena (recombination, mutation, and dispersal) prompted me to investigate the possible unification of these phenomena, which is provided in \citet{Altenberg:1984}.  

It turns out that the only way a parameter can enter linearly in the recursion is if it modifies transmission probabilities rather than fitnesses.  The approach to unification was to represent all of the models in one general expression, in which the specifics of the transmission probabilities $P(i \la j, k)$ (parents $j$ and $k$ produce offspring $i$) are ignored, while the \emph{variation} produced by the modifier locus is made explicit.  

The form of variation studied was where the modifier gene produced an equal scaling, $m$, of all transmission probabilities between states, \ie $ m P(i \la j, k)$, when $j \neq i$ or $k \neq i$.  The principle models that exhibited the reduction principle all incorporated this form of variation.  Equal scaling of transmission probabilities occurs when a single transformative event acts on the transmitted information, and the modifier gene controls the rate of this event \citep{Altenberg:2011:Mutation}.

With this explicit representation of variation,  the models that had exhibited the reduction principle had stability matrices of the form $\M(m) \D$ for newly introduced modifier alleles, where $\M(m) = (1-m)\I + m \P$ as in Karlin's theorem.  Once this structure is made evident, application of Karlin's Theorem 5.2 immediately yields the result that the growth rate of a new modifier allele was a decreasing function of $m$, so if it reduced $m$ below the current value in the population, it would invade, and if it increased $m$ above the current level, it would go extinct.  

Thus evolution would reduce the rates of all of these various processes, or others that had never been modeled before but which were covered by the general formulation.  Prior studies needed to assume only two alleles under selection, or two patches subdividing the population, because they relied on closed-form solutions to determinants or   eigenvalues.  Karlin's theorem allowed the result to be generalized to arbitrary numbers of alleles and patches, arbitrary patterns of transformation, and arbitrary selection regimes.

It should be noted that modifiers of segregation distortion have altogether different dynamics that merit a separate classification \citep[pp. 170--178]{Altenberg:1984}.

Slight variation among different models led to separate treatments for modifiers of mutation and recombination (\citealt[pp. 106--169]{Altenberg:1984}, \citealt{Altenberg:and:Feldman:1987}), modifiers of dispersal \citep[pp. 77--81, 178--199] {Altenberg:1984}, modifiers of rates of asexual vs. sexual reproduction ({ibid.} pp. 199--203), and culturally transmitted modifiers of cultural transmission --- i.e.\ `traditionalism' ({ibid.} pp. 203--206).
All of these phenotypes manifest the reduction principle for the same underlying reason, the spectral radius property shown in Karlin's Theorem 5.2.

\subsection{The Dispersal Modifier Results of Altenberg (1984)}

The results on the evolution of dispersal modifiers in \citet[pp. 77--81, 178--199]{Altenberg:1984} will be briefly reviewed, so that the work need not be duplicated, as has recently occurred \citep{Kirkland:Li:and:Schreiber:2006}. 

The model is of an organism that has a multiple-stage life cycle, consisting of random mating, semelparous reproduction, selection on gametes, zygotes, and adults, and lastly, dispersal.  The reproductive output of an organism depends on its patch and its diploid genotype for a gene under selection.  The probability of dispersing between any two patches is scaled by a modifier gene.  The model includes several generalizations of prior work: 
\itemlist{
\item arbitrary numbers of patches;
\item arbitrary dispersal patterns between patches, which may include cycles and asymmetry; 
\item dispersal of either adults or gametes (but not dispersal of zygotes, which breaks the Hardy-Weinberg  frequencies of diploids and complicates the analysis); 
\item arbitrary hard or soft selection patterns on diploids and gametes; 
\item arbitrary numbers of alleles at a dispersal-modifying locus; and 
\item arbitrary number of alleles for the locus with patch-specific fitnesses.
}

Analysis is made of the evolutionary stability of populations near equilibrium.  In order for any new modifier allele to grow or decline at a geometric rate, the equilibrium must 
possess variation in the reproductive rates among patches and/or genotypes.  This variation was first identified as a property of equilibrium populations at mutation-selection balance by \citet{Haldane:1937:Effect}, and was later called the `genetic load' by \citet{Muller:1950}.  

The term ``fitness load'' was used in \citet{Altenberg:1984} to generalize the genetic load concept to circumstances where there may be no genes involved --- in particular, to  patches with different growth rates where the stationary distribution leaves some patches as sinks and others as sources, as they were later to be called \citep{Pulliam:1988:Sources}.  The term `selection potential', $V \eqdef \max_i( D_i) / \text{mean}(D_i) - 1$, was adopted in \citet{Altenberg:and:Feldman:1987} because of the analogy to potentials in physical systems, and because $V$ was the actual maximum potential selective advantage that a modifier allele could accrue.  $V > 0$ is necessary for any geometric growth in the modifier allele.  The condition $V = 0$ corresponds to a population at an `ideal free distribution' \citep{Fretwell:and:Lucas:1969:Territorial,Fretwell:1972:Populations}.  

For the dispersal modifier model in \citet{Altenberg:1984}, a positive selection potential requires some differences at equilibrium among the terms
\an{\label{eq:MigrationSelectionPotential}
\frac{N^\sel(e) \ \wb(e,i)}{N ^\disp(e) \ \wb(e)},
}
over the environments $e$, and genotypes $i$, where
\desclist{
\item[$N^\sel(e)$] is the population size in environment $e$ after selection, and $N^\disp(e)$ after dispersal,
\item[$\wb(e)$] is the mean fitness in environment $e$, 
\item[$\wb(e, i)$] is the mean fitness of the allele $i$ under selection in environment $e$,
\item[$N^\sel(e) = N(e) \, \wb(e)$] under hard selection, and $N^\sel(e)$ is constant under soft selection.
}

One can see the two sources for a selection potential in \eqref{eq:MigrationSelectionPotential}:  ecological, i.e.\ variation in $N^\sel(e) / N^\disp(e) $ (mentioned in the earlier quote of \citealt{Olivieri:Michalakis:and:Gouyon:1995}), and genetic, i.e.\ variation in $\wb(e,i) / \wb(e)$.

Ideal free distributions having $V=0$ may be produced by the ``balanced mixture polymorphisms'' discussed in  \citet[pp. 101--104, 129, 189--190, 218--222] {Altenberg:1984}, which are synonymous with the Nash equilibria studied in \citet{Schreiber:and:Li:2011:Periodic}.

The main results obtained are the manifestation of the Reduction Principle for dispersal rates.  First, we have this result for modifier allele with extreme effect:
\begin{Result*}{3.27, \citet[p. 195]{Altenberg:1984}}
A modifier allele which stops all migration will always increase when introduced to a population with an equilibrium selection potential, for any linkage to the locus under selection.
\end {Result*}

For modifier alleles with intermediate effects on dispersal, tractability requires the assumption of tight linkage between the modifier locus and the selected locus.  Under tight linkage, the stability matrix for the new modifier allele becomes a direct sum of blocks for each allele $i$ under selection:
\an{\label{eq:MigrationEpsi}
\epsi_i(t+1) = \D_1 [ (1-m)\I + m \Mmb] \D_2(i) \ \epsi_i(t)
}
where $\Mmb$ is the matrix of average dispersal probabilities produced by modifier alleles in the equilibrium population, and
\ab{
\D_1 \!=\!\! \diag{\!\frac{1}{N^\disp(e)}\!}_{e=1}^{n_E}, 
\D_2(i) \!=\!\! \diag{\!\frac{N^\sel(e) \,  \wb(e, i)}{\wb(e)} \!}_{e=1}^{n_E},
}
where $n_E$ is the number of patches.  Then the following is obtained:
\begin{Result*}{3.28, \citet[p. 199]{Altenberg:1984}}
\enumlist{
\item The new modifier allele, ${\sf a}$, can change frequency at a geometric rate,
that is, $\rho(\M_{\sf a} \D_1 \D_2(i) ) \neq 1$, only if there is an equilibrium selection potential in the population, so that $\D_1 \D_2(i)  \neq \I$.
\item The spectral radius for the new modifier allele, ${\sf a}$, depends only on how
its marginal migration matrix $\M_{\sf a}$ is related to the equilibrium
marginal migration matrix $\Mmb$.  The results of Theorem 3.14 for
linear variation \ldots 
therefore apply directly:
}
\end {Result*}
\begin{Theorem*}{3.14, \citet[p. 137]{Altenberg:1984}}:
For a tightly linked modifier locus, when a new modifier allele, ${\sf a}$, is
introduced to a population at a stable viability-analogous,
tensor product equilibrium (VAHW), where there is a variance in the
marginal fitnesses of the selected types present, then for $m$ as defined in \eqref{eq:MigrationEpsi}, the new modifier allele frequency will increase if $m < 1$, and it will be excluded if $m > 1$.
\end{Theorem*}
Theorem 3.14 derives directly from Karlin's Theorem 5.2, which shows in addition that asymptotic growth rate of the new modifier allele increases as $m$ decreases throughout the range of $m$.

Karlin's Theorem 5.2, and the dispersal modifier results above, have recently been duplicated by \citet [Theorem 3.1]{Kirkland:Li:and:Schreiber:2006}.  They use a novel, structure-based proof for their version of Theorem 5.2, while Karlin used the \citeauthor{Donsker:and:Varadhan:1975} formula for the spectral radius.  They apply it to the evolution of unconditional dispersal, and prove a special case of \citet[Result 3.28 and Theorem 3.14]{Altenberg:1984} where the genetics and life history stages are absent.  Their results are extended to continuous time models by \citet[online Appendix B]{Schreiber:and:Lloyd-Smith:2009}, while \citet{Altenberg:2010:Karlin} uses the  \citeauthor{Donsker:and:Varadhan:1975} formula to extend Theorem 5.2 to the continuous time case.

The results in \citet{Kirkland:Li:and:Schreiber:2006}, while being special cases of \citet {Altenberg:1984} as far as the genetics are concerned, offer generalizations of the reduction principle in other new directions, namely, they generalize the work on density-dependent population regulation first addressed for dispersal modifiers by  \citet{Asmussen:1983:Evolution}, and cover the general case where growth rates decrease with population size \citep[Assumptions A1-A3]{Kirkland:Li:and:Schreiber:2006}.  They also cover the case of reducible dispersal matrices (Theorem 4.4), the case of lossy dispersal (Assumption A4), and the fate of the modifier allele far from perturbation (Theorem 3.3).  They examine conditional dispersers, and analyze the evolutionarily stable state in which dispersal has been conditioned to the point where the population reaches an ideal free distribution.

It should be noted that the ideal free distribution was proposed as ultimate evolutionarily stable state by \citet{Kimura:1967} in his `principle of minimum genetic load'.  Kimura was thinking about the evolution of mutation rates, not dispersal.  But the driving force in each case --- the genetic load for mutation, and the presence of sink and source populations for dispersal \citep{Pulliam:1988:Sources} --- is mathematically the same phenomenon.

\section{Departures from Reduction}

While the reduction phenomenon occurs throughout a diverse class of evolutionary models, there are two principal classes in which departures from reduction are found.  The first class, which will not be further addressed here, comprises situations in which the population is continually kept far from equilibrium, due to genetic drift (e.g. the Hill-Robertson effect \citep{Barton:and:Otto:2005:Recombination,Roze:and:Barton:2006:Hill-Robertson,Keightley:and:Otto:2006:Interference}, also \citet{Gillespie:1981:Mutation}), varying selection regimes \citep{Charlesworth:1976:Recombination,Gillespie:1981:RoleIII,Ishii:Matsuda:Iwasa:and:Sasaki:1989,Sasaki:and:Iwasa:1987,Bergman:and:Feldman:1990:More, Wiener:and:Tuljapurkar:1994,Schreiber:and:Li:2011:Periodic,Blanquart:and:Gandon:2010:Migration}, or flux of beneficial mutations \citep{Eshel:1973:Mutation,Eshel:1973:Modifying,Kessler:and:Levine:1998:Mutator}.  

The second class comprises cases of populations near equilibrium where multiple transformation processes act on the transmissible information of the organism.  Studies of multiple transformation processes where departures from reduction are found include the evolution of:
\itemlist{
\item \emph{recombination in the presence of mutation} \citep{Feldman:Christiansen:and:Brooks:1980,Charlesworth:1990:MSB,Otto:and:Feldman:1997,Pylkov:Zhivotovsky:and:Feldman:1998}.  The greatest attention has been given to this combination.  The departures from the reduction result in this case are the basis of the `deterministic mutation hypothesis' for the evolution of sex \citep{Kondrashov:1982,Kondrashov:1984,Kouyos:Silander:and:Bonhoeffer:2007:Epistasis}.  
\item \emph{recombination in the presence of dispersal} \citep{Charlesworth:and:Charlesworth:1979,Pylkov:Zhivotovsky:and:Feldman:1998}; 
\item \emph{multiple mutation processes} \citep[pp. 137--151]{Altenberg:1984}; 
\item \emph{recombination in the presence of  segregation and syngamy} (which self-fertilization exposes in the recursion) \citep{Charlesworth:Charlesworth:and:Strobeck:1979,Holsinger:and:Feldman:1983:LM}; 
\item \emph{mutation in the presence of segregation and syngamy} (exposed in the recursion by self-fertilization \citep{Holsinger:and:Feldman:1983:MM}, or fertility selection \citep{Holsinger:Feldman:and:Altenberg:1986,Twomey:and:Feldman:1990:Mutation}).  
}
It is notable that in their studies of dispersal in the presence of mutation, \citet{Wiener:and:Feldman:1991:Evolution,Wiener:and:Feldman:1993:Effects} found no departures from the reduction principle.

The pattern of departures from the reduction principle caused by multiple transformation processes was summarized in \citet[pp. 149, 225--228]{Altenberg:1984} by a simple heuristic:
\desclist{
\item[The principle of partial control:]  When the modifier gene has only partial control over the transformations occurring at loci under selection, then it may be possible for the part which it controls to evolve an increase in rates.  
}
In several cases where multiple transformation processes produce departures from reduction, the stability matrix on the modifier gene has the form
\an{\label{eq:PartialControl}
\M(m) = (1-m) \A + m \B.
}
Matrices of the form \eqref{eq:PartialControl} also appear when the modifier gene is not tightly linked to the loci under selection (\citealt{Feldman:1972}, \citealt[p. 135]{Altenberg:1984}, \citealt{Altenberg:and:Feldman:1987}, \citealt{Altenberg:2009:Linear}).  Karlin's Theorem 5.2 does not apply to such matrices, leaving an entire class of models as an unsolved open problem.  In a survey of open problems in the spectral analysis of evolutionary dynamics \citep{Altenberg:2004:Open}, the following problem was posed:
\begin{OpenQuestion}[3.1 in {\citealt{Altenberg:2004:Open} }]
Let $\A$ and $\B$ be irreducible stochastic matrices, and let $\D \neq c \, \I$ be a positive diagonal matrix.  Define
\an{\label{eq:OpenProblem}
\M(\mu ) = (1 - \mu )\A + \mu \B.
}
For what conditions on $\A$, $\B$, and $\D$ is the spectral radius $\rho(\M(\mu ) \D)$ strictly decreasing in $\mu$, for $ 0 \leq \mu \leq 1$, or
\ab{
\df{}{\mu} \rho(\M(\mu )\D) < 0 ?
}
\end{OpenQuestion}
This open problem brings us back to the paper by \citet{\McD}.

\section{The Model of McNamara and Dall (2011)}

The model of \citet[eq. (D.10)]{\McD} is an example of `partial control' \eqref{eq:PartialControl}, where we set $\A = \P$ and $\B = \piv \evt$, $\piv$ being the stationary distribution for stochastic matrix $\P$, i.e.\ $\P \piv = \piv$.  A major point of interest is that \McDA\  find conditions on $\P$ that produce departures from the reduction phenomenon, providing another example of the principle of partial control, and contributing towards answering the open problem posed in \citet{Altenberg:2004:Open}, above.  The recursion for their model is
\an{\label{eq:McDRecursion}
\z(t+1) = \M(m) \,  \D \, \z(t),
}
where
\an{\label{eq:McDgeneral}
\M(m) \eqdef  (1-m) \Pm + m \, \piv \evt .
}
The control exerted by $m$ over the transformations occurring in the system in \eqref{eq:McDgeneral} is only partial because the environment itself undergoes transformation, represented by $\P$, and the organism cannot eliminate $\P$, but only shift between $\P$ and $\piv\evt$.

The \McDA\ model represents the following.  Let $z_i(t)$ be the number of individuals in environment $i$ at time $t$, and $z_i(t+1)$ be the number after one iteration of reproduction and dispersal.
\enumlist{
\item An individual is born into a site with environment type $i$; 
\item The individual reproduces on the site, and produces an average of $D_i$ offspring when in environment $i$;  
\item Each offspring disperses independently with probability $m$ to a random site; 
\item In one generation, sites of environment type $j$ change randomly and independently to type $i$ with probability $M_{ij}$;  
\item The sites have settled down to a stationary distribution, so the probability that the site will be in environment state $i$ is $\pi_i$.  
}
Recursion \eqref{eq:McDRecursion} in summation form is:
\ab{
z_i(t+1) = (1-m)  \sum_j P_{ij} D_j z_j(t)  + m  \pi_i \sum_j D_j z_j(t).
}
\citet {\McD} obtain analytical results for the case of $n=2$ types of environment:
\ab{ 
\Bmatr{z_1(t+1)\\z_2(t+1)} &=  \M \D  \Bmatr{z_1(t) \\ z_2(t)},
\text{ where } \D = \Bmatr{D_1 \tb 0 \\ 0 \tb D_2},
}
and
\an{  \label{eq:McD2}
\M &= (1 \!- \!m)\!\! \Bmatr {1-P_{21} \tb P_{12}\\  P_{21} \tb 1-P_{21}} + m \!\Bmatr {\pi_1 \tb \pi_1  \\ 1-\pi_1 \tb  1-\pi_1 }.
}

The model is notable for how it represents environmental randomness.  The common way to model randomly changing environments would be to let $\z(t)$ represent the population size in each patch, $\M$ represent the dispersal between patches, and let the matrix of environment-specific growth rates, $\D$, be a random or time-dependent variable on each patch (e.g. \citealt[pp. 90--92, 103--104, 140--145]{Karlin:1982}), yielding a system 
\an{\label{eq:TimeVaryingD}
\z(t) = \M\, \D_{(t)}\, \M\, \D_{(t-1)} \ldots \M\, \D_{(2)}\, \M\, \D_{(1)}\, \z(0).
}
The analysis of such models can be challenging, requiring a resort to approximations and numerical analysis (see \citealt{Gillespie:1981:RoleIII}, \citealt{Tuljapurkar:1990:Population}, \citealt{Wiener:and:Tuljapurkar:1994}).  Progress is being made in this area, however, for example the analysis of a two-cycle model for the evolution of dispersal by \cite{Schreiber:and:Li:2011:Periodic}.

When the random process of changing environments is independent among all the patches, as the number of patches becomes large, the system becomes deterministic in the same way that the Wright-Fisher model becomes deterministic for large populations.  This allows one to stop keeping track of each \emph{patch}, and just keep track of the number of individuals in each \emph{environment} type, which is what \McDA\ do in \eqref {eq:McDRecursion}.  This tremendously simplifies the analysis.

\subsection{Clues to the Generalization of the Results}
The original motivation for this paper was to generalize the results of \citet{\McD} to an arbitrary number of environments, and to gain insight into why their model produces departures from the reduction phenomenon.  
Their results reveal four clues needed to solve this generalization:
\enumlist{
\item \emph{The harmonic mean}: 
\citet{\McD} find that departures from the reduction phenomenon are determined by the critical condition $\tau_1^{-1} + \tau_2^{-1} < 1$, where $\tau_i$ is the expected duration of environment $i$.  This expression is part of the harmonic mean, $2 / (\tau_1^{-1} + \tau_2^{-1})$.  Could the harmonic mean of $\tau_i$ figure into a generalization of their results?
\item \emph{The limiting distribution}: 
The two matrices in \eqref{eq:McDgeneral}, $\P$, and $\piv \evt$, are not arbitrary, but have the relation $\piv \evt = \lim_{t \goesto \infty} \P^t$.  This means, notably, that they commute:  $\P (\piv \evt) = (\piv \evt) \P = \piv \evt$, and thus satisfy one key condition of Karlin's Theorem 5.1.
\item \emph{The second eigenvalue}:  
The terms $\tau_1^{-1}$ and $\tau_2^{-1}$ derive from the probabilities in $\P$:  $\tau_1^{-1} = 1 - P_{11}$ and $\tau_2^{-1} = 1 - P_{22}$.  The condition $\tau_1^{-1} + \tau_2^{-1} < 1$ translates to $1 < P_{11} + P_{22}$.  Is it a coincidence that $P_{11} + P_{22}$ appears in the second eigenvalue of $\P$, $\lambda_2(\P) = 1 - P_{11} - P_{22}$?  Because we see that the critical condition becomes $\lambda_2(\P) < 0$, which is precisely when $\P$ no longer meets the condition of Karlin's Theorem 5.1 that it be symmetrizable to a positive definite matrix.  By extrapolation, if all the eigenvalues of $\P$ besides $1$ are  negative, could this be a general condition for departures from reduction? 
\item \emph{Symmetrizability}:
Since clues 2. and 3. show the relationship between the results of \McDA\ an Karlin's Theorem 5.2, and we note that irreducible $2 \times 2$ matrices are always symmetrizable, might we want to retain symmetrizability in $\P$ as we try to generalize the results to $n \times n$ matrices?
}

By following the last clue and constraining $\P$ to be symmetrizable as in \eqref{eq:KarlinFamily}, we shall find it tractable to generalize the results of \McDT, and we shall see that the conjectures prompted by the first and third  clues are true.  

Symmetrizable stochastic matrices are equivalent to the transition matrices of ergodic reversible Markov chains \citep[Lemma 2]{Altenberg:2011:Mutation}.  A Markov chain is reversible when the probability of cycles in one direction equals the probability of cycles in the opposite direction \citep[Theorem 4.7.1, p. 127]{Ross:1983}.  In nature, directional cycles of environmental change may be more the rule than the exception, however.  Whether cyclical environments would produce different results remains an open question.

We can step beyond the \McDA\ model and obtain a more general theorem for departures from reduction for the form $\M(m) = \P [(1-m) \I + m \Q]$, where $\P$ and $\Q$ satisfy \eqref{eq:KarlinFamily}.  This is provided in Theorem \ref{Theorem:Main} in \emph{Results}.  The theorem in \citet{Altenberg:2009:Mutation,Altenberg:2011:Mutation} that generalizes the reduction principle to the evolution of \emph{mutation} rates among multiple  loci turns out to be a special case of Theorem \ref{Theorem:Main}.  This again illustrates the fact that genetic, spatial, cultural, and other transmissible information all belong to a single mathematical framework, and that results from one domain can often translate easily into results in other domains.

\section{Results}

\McDT\  describe their concept of a ``multiplier effect'' without ever giving it a precise mathematical definition.  But it is clear from their usage in \citet[online Appendix A, Theorem A]{\McD} that what they are thinking about can be summarized as the covariance between 1) the growth rates in each environment, and 2) the excess abundance of the population in that environment over what it would be without differential growth rates.  This is defined explicitly below as \emph{the fitness-abundance covariance}.  

When organisms are semelparous, and generations are discrete and non-overlapping, there are two phases in the life cycle that one can census the population: before and after dispersal, or equivalently, after and before reproduction.  Thus, the fitness-abundance covariance must be defined for both census phases.

The \fac\ is an object of interest in its own right.  Section \ref{section:FAC} ventures beyond the specifics of the \McDA\ model and explores various properties of the \fac\ for the completely general case of $\z(t+1) = \M \D \z(t)$, where $\M$ is a stochastic matrix representing any process of change between states, and $\D$ represents the state-specific growth rates.  The generality of results in Section \ref{section:FAC} not only includes the \McDA\ model as a special case,  
but goes beyond models of dispersal since $\M$ can just as well represent a mutation matrix between genotypes whose fitnesses are $D_i$.  The results can also apply to a rare genotype in a sexual population where $\M\D$ represents the linear stability matrix on its growth.

Section \ref{section:McD} returns to the specific model of \McDA\ with the chief goal of generalizing the results to any number of environmental states.  Here is where we pursue the clues described in the previous section.

\newcommand{\TableOne}{
\begin{table}[top]
\framebox[\columnwidth]{
\parbox[b]{3.2in}
{
\caption{Definitions and Symbols\label{Table:Definitions}}
{\small 
\desclist{
\item [$\A, \M, \D, \P, \S$] or other boldface capital letters represent $n \times n$ matrices, and $\vv, \x, \y, \ev$, or other bold face lower case characters represent $n$-vectors;  the identity matrix is $\I$; a scalar matrix is $c \, \I$ for $c \in \Reals$;
\item [ \pr{$A_{ij} \equiv [\A]_{ij}$}] represents the elements of $\A$,  $i,j = 1, \ldots, n$, and $x_i$ represents the elements of $\x$;
\item [ \pr{$D_i \equiv [D]_{ii}$}] represents the diagonal elements of diagonal matrix $\D$;
\item [{\it a positive diagonal matrix}] has $D_i > 0$, $i = 1, \ldots, n$;
\item [$\protect{[\A]^i}$] represents the $i$th row of matrix $\A$, and $[\A]_j$ represents the $j$th column.
\item [$\ev$] represents the {unit vector}, where all elements are $1$;
\item [$\ev_j$] represents the {$j$th basis vector}, which has $1$ at position $j$ and $0$ elsewhere;
\item [$\!\! \diag{\x} \equiv \D_\x$] is a diagonal matrix of the vector $\x$;
\item [$\A\tr$, $\z\tr$, $\evt,$] etc., represent the {transpose};
\item [$\lambda_i(\A) \equiv \lambda_{Ai},$] $i = 1 \ldots n$ represent the {eigenvalues} of $\A$;
\item [{\it symmetrizable to $\S$}] means that an $n \times n$ matrix can be represented as a product $\A = \Lm \Sm \R$, where $\Sm$ is a symmetric real matrix, and $\Lm$ and $\R$ are positive diagonal matrices; 
\item [{\it stochastic}] means an $n \times n$ matrix with nonnegative elements and whose columns (by convention here) sum to one ({column stochastic});
\item [{\it positive definite}] means a matrix that is symmetric and has only positive eigenvalues;
\item [{\it irreducible}] means an $n \times n$ nonnegative matrix where for every $i,j$ there is some $t$ such that $[\A^t]_{ij} > 0$;
\item [$\rho(\A) \eqdef \max_i | \lambda_i(\A) |$]  represents {the spectral radius,} the largest modulus of any eigenvalue of $\A$. 
\item [$\lambda_1(\A)$] by convention will refer to the {\it Perron root} of a nonnegative irreducible matrix $\A$, which is the positive eigenvalue guaranteed by Perron-Frobenius theory \citep[Theorems 1.1, 1.5]{Seneta:2006} to exist, to be the spectral radius, and to be as large as the modulus (i.e.\ magnitude) of any other eigenvalue.  So $\lambda_1(\A) = \rho(\A) \geq | \lambda_i(\A)|$ for $i = 2, \ldots, n$.
\item [$\vv(\A)$] and  $\uv(\A)\tr$ represent the right and left {\it Perron vectors} of nonnegative irreducible $\A$, the eigenvectors associated with the Perron root, guaranteed by Perron-Frobenius theory to be strictly positive.  So $\A \, \vv(\A) = \rho(\A) \, \vv(\A)$, and $\uv(\A)\tr \A = \rho(\A)\,  \uv(\A)\tr$.  By convention $\evt \vv(\A) = 1$ and $\uv(\A)\tr \vv(\A) = 1 $.
\item[$\vv \equiv \vv(\A),$] $\uv \equiv \uv(\A)$, and $\rho \equiv \rho(\A)$, throughout, where $\A$ is obvious from context.
\item [$\piv \equiv \vv(\P)$] traditionally represents the stationary distribution of irreducible (column) stochastic matrix $\P$.
\item [{\it The harmonic mean}] of a set of numbers $\{ \tau_i \}$ is
\ab{
E_H(\tau_i) \eqdef \frac{1}{\displaystyle \frac{1}{n} \sum_{i=1}^n \dspfrac{1}{\tau_i} } \ .
}
}
}
}
}
\end{table}	
}
For clarity, terminology and conventions are provided in Table \ref{Table:Definitions}.

\subsection{The Fitness-Abundance Covariance}\label{section:FAC}

A precise definition needs to be given for the degree to which ``individuals tend to be in locations to which they are well suited.''   While it may sound reasonable that an organism's ``mere existence informs an organism that it is liable to be in favourable circumstances'' \citep[p. 237]{\McD}, this is not generally true.  

The following is an example where an organism is more likely to find itself in a sink habitat than a source habitat.  The situation is where there is a small source patch within a large habitat of sink patches.  We get a simple result if we assume the extremes:  that sink habitats are lethal, and dispersing organisms recruit with fixed probabilities $\pi_i$ to each patch, $i$, and the dispersal rate is $m$.  Then $v_1 = 1 - m(1- \pi_1)$ is the stationary proportion of the population in the source patch after dispersal.  The stationary portion of the population in the source patch, $v_1$, can be made as small as one wishes by large dispersal rate $m$, and small $\pi_1$.  

Clearly, \emph{before} dispersal, all organisms  in this example are in the source patch.  So one must be clear about when in the life cycle one is speaking.  An organism `deciding' on whether to disperse or not is obviously at the pre-dispersal phase.  But the post-dispersal phase is the phase that \McDA\ use to measure the `multiplier effect'.

Examination of the results of \McDA\ also reveals that when they speak of an organism being ``liable to be in favourable circumstances,'' what they actually mean is that an organism is \emph{more likely} to be in a favorable habitat than it \emph{would be} if there were no growth advantage there, not that the organisms is actually \emph{liable to be there}.  This is the concept that I make precise as the \emph{fitness-abundance covariance}.  Even in this relative value of abundance, however, we will see that the \fac\ is not always positive.

The \fac\ relates three different sets of values: the environment-specific growth rates $D_i$, the stationary distribution in the \emph{presence} of differential growth rates, referred to as $v_i$, and the stationary distribution in the \emph{absence} of differential growth rates, referred to as $\pi_i$.

The stationary distribution for recursion $\z(t+1) = \M \D \, \z(t)$ satisfies
\ab{
\rho \ \vv = \M \D \vv,
}
where $\vv$ is the eigenvector of $\M\D$ associated with the largest eigenvalue of $\M\D$, $\rho$.  This is called the right Perron vector.  Throughout, $\vv(\A)$ will represent the right Perron vector of a matrix $\A$ (see Table \ref{Table:Definitions}).  \TableOne

The magnitude of $\rho$ determines whether the population grows ($\rho>1$) or declines ($\rho<1$) or is stationary ($\rho=1$), and ecological models typically impose some kind of negative density dependence so that as population size $\z$ gets large enough, $\rho$ decreases with $\z$, and a stationary state of $\rho=1$ can be attained.  The problems addressed here do not concern the absolute value of $\rho$, but only the relative changes to $\rho$ and $\vv$ under changes in $\M$ and $\D$.  For a general treatment of negative density dependence, \citet{Kirkland:Li:and:Schreiber:2006} provide a thorough analysis.

The stationary distribution in the presence if differential growth rates depends on the phase in the life cycle at which the census is taken.  The life cycle consists of alternation between differential growth and dispersal, $\ldots \D \M \D \M \D \M \ldots$.   When censused just after dispersal, the stationary distribution is $\vv(\M\D)$.  Censused just before dispersal it is $\vv(\D\M)$.  

We see that $\vv(\M\D)$ and $\vv(\D\M)$ have a simple relationship from the cyclical structure.  $\D \,\vv(\M\D)$ is the Perron vector of $\D \M$ up to scaling, since
\ab{
\D [ \M \D \, \vv(\M\D) ] &= \rho(\M\D)\, \D \, \vv(\M\D)
}
When scaled to satisfy $\evt \vv(\D\M) = 1$, one gets the relationship: 
\an{\label{eq:vDM}
 \vv(\D \M)  =  \frac{1}{\rho(\M\D)} \D& \, \vv(\M\D).  
}
It should be noted that in continuous-time models such as quasispecies \citep{Eigen:and:Schuster:1977}, selection and transformation happen simultaneously so there are no separate life cycle phases, hence no distinction between pre- and post-dispersal stationary states.

For semelparous organisms with discrete, non-overlapping generations, the fitness-abundance covariance is now defined for both phases of the life cycle.
\begin{Definition*}[Fitness-Abundance Covariance]
\ 

The fitness-abundance covariance is defined as the unweighted covariance between the environment-specific growth rates and the excess of the stationary distribution above the distribution that the population would attain in the absence of differential growth rates:
\enumlist{
\item Post-dispersal:
\ab{
\FAC (\M \D) &\eqdef \Cov(D_i, v_i(\M\D) - v_i(\M)) & \\
&=\ \frac{1}{n} \sum_{i=1}^n D_i(v_i(\M\D) - v_i(\M))& \notag\\
&  \  \ -  \frac{1}{n} \sum_{i=1}^n D_i \ \frac{1}{n}\sum_{j=1}^n  (v_j(\M\D) - v_j(\M)). & \notag
}
\item Pre-dispersal:
\ab{
\FAC (\D \M ) &\eqdef  \ \Cov(D_i, v_i(\D\M) - v_i(\M)) &\\
&= \  \frac{1}{n} \sum_{i=1}^n D_i(v_i(\D\M) - v_i(\M)) &\notag \\
&\ \  -  \frac{1}{n} \sum_{i=1}^n D_i  \ \frac{1}{n} \sum_{i=1}^n  (v_i(\D\M) - v_i(\M)). & \notag
}
}	
\end{Definition*}
Several elementary results are described.  The first shows that the relationship between the pre- and post-dispersal fitness-abundance covariances is Fisher's Fundamental Theorem of Natural Selection in a slightly new context.

\begin{Theorem-InOrder}[Fitness-Abundance Covariance and Census Phases]\label{Theorem:FACFisher}
\ 

Let $\M$ be an irreducible column stochastic matrix and $\D$ a positive diagonal matrix.

Then
\ab{
\FAC (\D \M ) = \FAC (\M \D ) + \frac{1}{n \, \rho(\M\D)} \Var_{\vv}(D_i),
}
where $\Var_{\vv}(D_i)$ is the $\vv(\M\D)$-weighted variance of $D_i$,
\ab{
\Var_{\vv}(D_i) \eqdef \sum_{i=1}^n v_i(\M\D) D_i^2 - \left(\sum_{i=1}^n v_i(\M\D) D_i\right)^2  .
}
\end{Theorem-InOrder}
\begin{proof}
Here, $\piv \equiv \vv(\M)$, $\vv \equiv \vv(\M\D)$, and $\rho \equiv \rho(\M\D)$.  
We first note that
\ab{
\sum_{i=1}^n D_i v_i =  \evt \D \, \vv &= \evt \M \D \, \vv = \rho \evt \vv = \rho.
}
\an{
\FAC(\M\D) &= \Cov(D_i, v_i - \pi_i) \notag \\& 
= \frac{1}{n} \sum_{i=1}^n D_i (v_i - \pi_i ) - 
\frac{1}{n} \sum_{i=1}^n D_i  \frac{1}{n} \sum_{j=1}^n  (v_j - \pi_j ) \notag\\
&= \frac{1}{n} \left( \rho - \sum_{i=1}^n D_i \pi_i   \! \right)  - 0 \label{eq:CovMD}.
}
Substitution with \eqref{eq:vDM} and \eqref{eq:CovMD} gives
\an{
&\FAC (\D \M ) = \ \Cov(D_i, v_i(\D\M) - \pi_i) \notag \\
&=   \frac{1}{n} \sum_{i=1}^n D_i(\frac{1}{\rho} D_i v_i - \pi_i) 
-  \frac{1}{n^2} \sum_{i=1}^n D_i \sum_{j=1}^n  (v_j(\D\M) - \pi_j)  \notag \\
&=  \frac{1}{n }\left(\frac{1}{\rho} \sum_{i=1}^n D_i^2 v_i 
-  \sum_{i=1}^n D_i \pi_i\right)  - 0 \label{eq:FACDM} \\
&=  \frac{1}{n }\left(\frac{1}{\rho}\left[\sum_{i=1}^n D_i^2 v_i - (\sum_{i=1}^n D_i v_i)^2 \right]
+ \rho -  \sum_{i=1}^n D_i \pi_i \! \right) \label{eq:FACDMVar} \\
&=  \frac{1}{n \, \rho } \Var_\vv(D_i) + \FAC(\M\D). \notag \qedhere
}
\end{proof}
\begin{Corollary-InOrder}[Derivatives of Fitness-Abundance Covariances and $\rho$]\label{Corollary:rhoFACderiv}
\ 

Let $\M(m)$ be a family of irreducible stochastic matrices, differentiable in $m$, and assume $\vv(\M(m)) = \piv$ for all $m \in (0, 1]$.  Let $\D$ be a positive diagonal matrix.   Set $\rho \equiv \rho(\M(m)\D)$.

Then
\an{\label{eq:CovDv}
\df{}{m}&\FAC(\M(m)\D) 
= \frac{1}{n} \df{\rho}{m},
\intertext{and}
\df{}{m}& \FAC(\D\M(m)) \notag \\
=& \frac{1}{n} \left( \df{\rho}{m}  + \frac{1}{\rho} \df{}{m} \Var_\vv(D_i)  - \frac{1}{\rho^2} \Var_\vv(D_i) \right)\label{eq:FACDMderiv}.
}
\end{Corollary-InOrder}
\begin{proof}
Differentiation of \eqref{eq:CovMD} and \eqref {eq:FACDMVar} directly gives \eqref{eq:CovDv} and \eqref {eq:FACDMderiv}.
\end{proof}

By plain intuition we would expect the \fac\ to be positive.  But \McDA\ found circumstances in which the fitness-abundance covariance is negative just after dispersal.  The question then remains, what about just before dispersal, when individuals are still in the environment whose growth rate they just replicated under?  Here is where intuition suggests the \fac\ should be positive.  This is proven to be the case when $\M$ is the transition matrix of a reversible Markov chain with positive eigenvalues.  But a counterexample is provided when $\M$ represents a periodic chain that cycles through the states, which has complex eigenvalues.

\begin{Theorem-InOrder}[Positivity of the Pre-Dispersal Fitness-Abundance Covariance]\label{Theorem:PositiveFACDM}
\ 

Let $\M$ be the transition matrix of an ergodic reversible Markov chain, with only nonnegative eigenvalues.  Let $\D \neq c \, \I$ be a positive diagonal matrix.

Then
\ab{
\FAC (\D \M ) \eqdef  \ \Cov(D_i, v_i(\D\M) - v_i(\M)) > 0.
}
\end{Theorem-InOrder}
\begin{proof}
Here, $\piv \equiv \vv(\M)$, $\vv \equiv \vv(\M\D)$, and $\rho \equiv \rho(\M\D)$.  From \eqref {eq:FACDM},
\ab{
\FAC(\D\M) &> 0 \iff
\sum_{i=1}^n D_i^2 v_i >  \rho\sum_{i=1}^n D_i \pi_i.
}
Since $\D \neq \c \, \I$, and $\vv > \0$,
\an{\label{eq:VarD}
\Var_\vv(D_i) = \sum_{i=1}^n D_i^2 v_i - (\sum_{i=1}^n D_i v_i)^2 > 0.
}
The condition that $\M$ be the transition matrix of an ergodic reversible Markov chain is equivalent to it being diagonally similar to a symmetric matrix (\citealt[Proposition 1.3B]{Keilson:1979}; \citealt[Lemma 2]{Altenberg:2011:Mutation}).  Since $\M$ has all nonnegative eigenvalues, that matrix is positive semidefinite.  This allows application of the inequality in \citet[Theorem 4.1]{Friedland:and:Karlin:1975}:   $\rho(\D\M) \geq \sum_{i=1}^n D_i \pi_i$.  In \eqref {eq:VarD} this gives
\ab{
 \sum_{i=1}^n D_i^2 v_i > (\sum_{i=1}^n D_i v_i)^2 = \rho^2 \geq \rho \sum_{i=1}^n D_i \pi_i. &\qedhere
}
\end{proof}

For a counterexample to the positivity of the pre-dispersal \fac, we try a transition matrix $\M$ that is as far from Theorem \ref {Theorem:PositiveFACDM} as possible, so the states are periodic and the  eigenvalues other than 1 are complex roots of unity.  This represents the situation of pelagic organisms along a gyre (e.g. \citealt{Cowen:Pari:and:Srinivasan:2006:Scaling}).

\newcommand{\cdotssm}{{\cdots}}
\newcommand{\vdotssm}{{\vdots}}
\newcommand{\ddotssm}{{\ddots}}

\begin{Theorem-InOrder}\label{Theorem:CyclicMD}
Let $\M$ be an $n$-cyclic matrix,
\ab{
\M = \Bmatr{
0  & 0  & 0  & \cdotssm & 0 & 1\\ 
1  & 0  & 0  & \cdotssm & 0 & 0\\ 
0  & 1  & 0  & \cdotssm & 0 & 0\\ 
&   & 1  & \ddotssm & \vdotssm & 0\\ 
\vdotssm &  &   & \ddotssm & 0 & \vdotssm \\ 
0  & 0  & 0  & \cdotssm & 1 & 0 \\ 
}.
}

Then for $\D \neq c \, \I$,  
\an{\label{eq:Cyclic}
\FAC(\M\D) 
&=
\frac{1}{n} \left(   \prod_{i=1}^n D_i^{1/n} - \frac{1}{n} \sum_{i=1}^n D_i    \right)   < 0
}

\end {Theorem-InOrder}
\begin{proof}
For a matrix cyclic in this direction, given any $\z>0$, $[\M\D \z]_{(i \text{ mod } n) +1} = D_i z_i$.  Thus $(\M\D)^n \z = ( \prod_{i=1}^n D_i ) \z$.  So $(\M\D)^n \vv(\M\D) = ( \prod_{i=1}^n D_i ) \vv(\M\D) = \rho(\M\D)^n \vv(\M\D)$.  Hence
$\rho(\M\D) = \prod_{i=1}^n D_i^{1/n}$.  Substitution in \eqref {eq:CovMD} gives \eqref{eq:Cyclic}.
$\FAC(\M\D)$ is negative because it is $1/n$ times the difference between the geometric and arithmetic means of $D_i$, which is always negative if not all $D_i$ are equal \citep[pp. 20--26]{Steele:2004:Cauchy}.
\end{proof}

\begin{Theorem-InOrder}\label{Theorem:CycleCounterexample}
When the states are transformed in a cycle, it is possible for the pre-dispersal fitness-abundance covariance to be negative.
\end{Theorem-InOrder}
\begin{proof}
An example is constructed.  Let $\M$ represent the period-3 cycle of states $1 \goesto 2 \goesto 3 \goesto 1 \ldots$
\ab{
\M = \Bmatr{0  & 0 & 1\\ 1 & 0 & 0 \\ 0 & 1 & 0},
}
and let $\D = \diag{D_1, D_2, D_3}$. 
 
The spectral radius is $\rho = (D_1 D_2 D_3)^{1/3}$.  By symmetry, $\pi_i = 1/3$, $i = 1, 2,3$.  Symbolic computation with \emph{Mathematica\texttrademark} shows that
\ab{
&\vv(\D\M) \\ 
&= \Bmatr{ \left(1 + D_2^{2/3}/(D_1^{1/3} D_3^{1/3}) + (D_2^{1/3} D_3^{1/3})/D_1^{2/3} \right)^{-1} \\
 \left( 1 + (D_1^{1/3} D_3^{1/3})/D_2^{2/3} + D_3^{2/3}/(D_1^{1/3} D_2^{1/3}) \right)^{-1} \\
 \left( 1 + (D_1^{1/3} D_2^{1/3})/D_3^{2/3} + D_1^{2/3}/(D_2^{1/3} D_3^{1/3}) \right)^{-1}}. &
}

A numerical survey shows that $\FAC(\D\M)$ is positive over most values of $(D_1, D_2, D_3)$ except for a very narrow range of $\D$ near the boundary where $\FAC(\D\M)$ becomes negative.  One such value is $(D_1, D_2, D_3)=({1/40000, 1, 1/8})$, 
which yields $\rho=0.0146$, 
$\vv(\M\D)\tr= (0.894, 0.0015, 0.105)$, 
$\vv(\D\M)\tr = (0.0015, 0.105, 0.894)$, and 
$\FAC(\D\M) = - 0.159$.
\end{proof}
The point of this odd counterexample is not that it represents something we might find in nature, but rather to say that we cannot entirely trust our intuition about the \fac, and that something more subtle is going on mathematically than we might suppose.

\subsection{Individual Stationary State Frequencies}
Let us now examine the relationships between individual values of $v_i$, $D_i$ and $\pi_i$.  

\McDA\ show for the case of $n=2$ that the post-dispersal fitness-abundance covariance is positive or negative depending on the durations of the environments, restated in their terms here:
\begin{Theorem-InOrder}[{\citet[online Appendix A, Theorem A]{\McD}}]  
\label{Theorem:AMcD}
Let $n=2$ in \eqref{eq:McDgeneral} to give \eqref{eq:McD2}. 
\enumlist{
\item If $\tau_1^{-1} + \tau_2^{-1} < 1$ then $\rho (\M \D) > \sum_{i=1}^2 D_i \pi_i$ and
\enumlist{
\item $D_1 < D_2 \implies v_2(\M \D) > v_2(\M) = \pi_2$
\item $D_1 > D_2 \implies v_2(\M \D) < v_2(\M) = \pi_2$.
}
\item If $\tau_1^{-1} + \tau_2^{-1} = 1$ then $v_i(\M \D) = v_i(\M) = \pi_i$, $i= 1, 2$, and $\rho (\M \D) =\sum_{i=1}^2 D_i \pi_i$.

\item If $\tau_1^{-1} + \tau_2^{-1} > 1$ then $\rho (\M \D) < \sum_{i=1}^2 D_i \pi_i$ and
\enumlist{
\item $D_1 < D_2 \implies v_2(\M \D) < v_2(\M) = \pi_2$
\item $D_1 > D_2 \implies v_2(\M \D) > v_2(\M) = \pi_2$.
}
}
\end{Theorem-InOrder}

The third case exhibits the very counterintuitive behavior that increasing the reproductive output of an environment will \emph{lower} the stationary proportion in that environment.  We can compare this result to the following general theorem on how changes to a matrix affect its Perron vector:

\begin{Theorem-InOrder}[{\citet[Theorem 2.1]{Elsner:Johnson:and:Neumann:1982}}] \label{Theorem:ElsnerEtAl}
\ 

Let $\A$ be an $n \times n$ nonnegative irreducible matrix.  Then for any nonnegative $n$-vector $\av \geq\neq \0$, $i \neq j \in \{1, \ldots, n\}$,
\an{\label{eq:Elsner}
\frac{v_i(\A + \ev_i \av\tr)}{v_i(\A)} > \frac{v_j(\A + \ev_i \av\tr)}{v_j(\A)}.
}
\end{Theorem-InOrder}
It is more useful for us to put it in the following form:
\begin{Corollary-InOrder}[Change in the Perron Vector] \label{Corollary:PerronVector}
\ 

When normalized to frequencies, $\evt \vv = 1$, then ${v_i(\A + \ev_i \av\tr)} > {v_i(\A)}  $.
\end{Corollary-InOrder}
\begin{proof}
The result follows immediately from rearrangement of \eqref{eq:Elsner} and summation using $\evt \vv = 1$:
\ab{
\sum_{ j \neq i}& \frac {v_j(\A + \ev_i \av\tr)}{v_i(\A + \ev_i \av\tr)} 
= \frac {1 - v_i(\A + \ev_i \av\tr)}{v_i(\A + \ev_i \av\tr)} \\
& < 
\sum_{ j \neq i} \frac{v_j(\A)}{v_i(\A)}
= \frac{1- v_i(\A)}{v_i(\A)} . \qedhere
}
\end{proof}

In this case, the behavior of the Perron root follows our intuition that increasing the $i$th row of $\A$ should increase the stationary proportion of $v_i$.

Something must be very different, therefore, between theorems \ref{Theorem:AMcD} and \ref{Theorem:ElsnerEtAl}, since they both deal with changes in the Perron vector when elements of the matrix are changed.  Theorem \ref{Theorem:AMcD} produces counterintuitive results that depend on $\tau_1^{-1} + \tau_2^{-1}$, while Theorem \ref{Theorem:ElsnerEtAl} has no conditions on details of the matrix.  How can this discrepancy be reconciled?  

We must write $\A$ in terms of $\M$ and $\D$ to compare the two results.  
Let $[\M]^i$ be the $i$th row of $\M$.   We can write
\ab{
\A + \ev_i \av\tr &= 
\Bmatr{
D_1 & 0 & 0 & 0 & 0& 0\\
0 & D_2 & 0 & 0 & 0 & 0\\
0& 0 & \ddots & 0 & 0 & 0\\
0 & 0 & 0 & D_i + \epsilon & 0 & 0\\
0& 0  & 0 & 0 & \ddots & 0\\
0 & 0 & 0 & 0 & 0& D_n \\
}
\M 
}
where $\A = \D \M$ and $\av = \epsilon [\M]^i$, $\epsilon > 0$.  Corollary \ref{Corollary:PerronVector} shows that increasing the reproductive output of environment $i$ from $D_i$ to $D_i + \epsilon $ increases the stationary proportion in environment $i$.  In the limit $\epsilon \goesto 0$, this gives:
\begin {Corollary-InOrder}\label{Corollary:dvDMdm}
For irreducible column stochastic matrix $\M$ and positive diagonal matrix $\D$:
\an{\label{eq:partialDM}
\pf{v_ \kappa(\D \M)}{D_\kappa} \geq 0. 
}
\end{Corollary-InOrder}

In the case $n=2$, we have $D_2 > D_1 \iff D_2 = D_1 + \epsilon$, $\epsilon > 0$, $\av\tr = \epsilon (M_{21}, M_{22})$, and
\ab{
\A + \ev_2 \av\tr &=  \Bmatr{D_1 & 0 \\ 0 & D_1+\epsilon} \M = \D \M.
}
So  Theorem \ref{Theorem:ElsnerEtAl} gives $v_2(\D\M) > v_2(\M)$ regardless of any details of $\M$.

The discrepancy is resolved by noticing that the order of $\M$ and $\D$ is reversed between Theorem \ref{Theorem:AMcD} and Theorem \ref{Theorem:ElsnerEtAl}.   The difference between the two is essentially in the phase of the life cycle at which the population is censused.  

We can contrast \eqref {eq:partialDM} with the following:
\begin{Corollary-InOrder}\label{Corollary:log_rho}
For irreducible column stochastic matrix $\M$ and positive diagonal matrix $\D$,
\an{
\pf{}{D_\kappa}\log(v_\kappa(\M\D)) \geq \pf{}{D_\kappa}\log\left(\frac{\rho(\M\D)}{D_\kappa}\right).\label{eq:partialLogv}
}
\end{Corollary-InOrder}
\begin{proof}
Substitution of \eqref{eq:vDM} in \eqref{eq:partialDM} and differentiation gives:
\ab{
&0 \leq \  \pf {v_\kappa(\D\M)}{D_\kappa}  
= \pf{}{D_\kappa}  \left( \dspfrac{D_\kappa v_\kappa (\M\D)}{\rho (\M\D)} \right) \\
&\ \ \ \ =  \frac{-1}{\rho^2}\pf{\rho}{D_\kappa} D_\kappa v_\kappa
+  \frac{1}{\rho} v_\kappa
+  \frac{1}{\rho}  D_\kappa \, \pf{v_\kappa}{D_\kappa} \iff \\
&0 \leq  \frac{-1}{\rho}\pf{\rho}{D_\kappa} 
+  \frac{1}{D_\kappa } +  \frac{1}{v_\kappa} \pf{v_\kappa}{D_\kappa} \iff \\ 
& \pf{}{D_\kappa}(\log(\rho)- \log(D_\kappa))
\leq \pf{}{D_\kappa}\log(v_\kappa) . \qedhere
}
\end{proof}

Comparing \eqref {eq:partialDM} and \eqref {eq:partialLogv} we see that the stationary distributions at different census phases behave differently.  

To summarize:
\itemlist{
\item the portion in environment $i$ censused before dispersal always increases with growth rate $D_i$; 
\item the portion in environment $i$ censused after dispersal can, under the right environment transition matrix, \emph{decrease} with increasing growth rate $D_i$.
}
This allows us to make inferences on the duration of the environments based on changes in the proportions of the population in each environment before and after reproduction:
\begin{Corollary-InOrder}[Census Inference on Durations of Environments]\label{Corollary:McDPhaseTau}
\ 

Consider the model of \McDA, $\z(t+1) = \M(m) \D \z(t)$, with $\M(m) =  [ (1-m) \Pm + m \, \piv \evt ]$ \eqref{eq:McDgeneral}, where $\piv = \P \piv = \vv(\P)$.  At a stationary distribution, let $\vv \equiv \vv(\M\D)$ be the vector of proportions of individuals in each environment before reproduction, and $\vv^\rec \equiv \vv(\D\M)$ be the proportions after reproduction.  For $n=2$ environments,
\enumlist{
\item If $v_1^\rec > \pi_1$, we know that $D_1 > D_2$.
\item If in addition, $v_1 <\pi_1$, then we know $E_H(\tau_1, \tau_2) < 2$; 
or if $v_1 > \pi_1$, then $E_H(\tau_1, \tau_2) > 2$.
}
\end{Corollary-InOrder}

\subsection{Generalization of The McNamara and Dall Model}\label{section:McD}
We shift now from general $\M$ to the specific model of dispersal in randomly changing environments of \McDT.  First, we see how the direction of selection on unconditional dispersal corresponds to the sign of the post-dispersal \fac.
\begin{Corollary-InOrder}[\McDA\ Model with general $n$]\label{Corollary:McDFACderiv}
\ 

Let $\M(m) \eqdef [ (1-m) \Pm + m \, \piv \evt ]$, where $\P$ is an irreducible stochastic matrix, and $\Pm \piv = \piv$.  Let $\D \neq c \, \I$ be a positive diagonal matrix.  Set $\vv\equiv \vv(\M\D)$.  Then
\an{
\FAC(\M\D) = \Cov(D_i, v_i - \pi_i) &> 0 \notag \\
\iff \df{}{m}\rho(\M(m)\D) &< 0, \label{eq:Cov>}
\intertext{(the reduction phenomenon) and}
\FAC(\M\D) = \Cov(D_i, v_i - \pi_i) &< 0 \\
\iff \df{}{m}\rho(\M(m)\D) &> 0. \label{eq:Cov<}
}
(departure from reduction).
\end{Corollary-InOrder}
\begin{proof}
At $m=1$, 
\ab{
\M(1) \D \ \vv =  \piv \evt \D \ \vv = \rho(\M(1) \D) \: \piv,
}
so $\vv = \piv$, hence $\Cov(\D_i, v_i(\M(1)\D) - \pi_i) = 0$.  For $m < 1$, \eqref{eq:Cov>} and \eqref{eq:Cov<} follow from \eqref{eq:CovDv}.
\end{proof}
{\bf Remark}.  It should be noted that this correspondence between the reduction phenomenon and the sign of the post-dispersal fitness-abundance covariance is specific to \McDA's model, $\M(m) = (1-m) \P + m  \vv(\P) \evt$.  Shortly we will examine the more general $\M(m) = \P[(1-m) \I + m  \Q]$, in which $\vv(\P\Q\D) \neq \vv(\P)$ generically, so $\Cov(\D_i, v_i(\M(1)\D) - \vv_i(\P)) \neq 0$.  Thus departures from reduction do not necessarily correspond to a negative fitness-abundance covariance.  For the general open problem $\M(m) = (1-m)\A + m \B$ \eqref{eq:OpenProblem}, it is not at all generic for $\vv(\A\D) = \vv(\A)$ or $\vv(\B\D) = \vv(\B)$, hence there is no general relationship between the  reduction phenomenon and the sign of the fitness-abundance covariance.

Next, the expression in \citet{\McD} involving the durations of the environments, $\tau_1^{-1} + \tau_2^{-1}$,  is generalized to the harmonic mean of the durations of $n$ environments.  The harmonic mean of the expected durations of states in a Markov chain (expected run lengths or `exit times') is shown to have a fundamental relationship to the sum of the eigenvalues (the trace) of its transition matrix, an identity whose earliest reference I find is \citet[p. 1017]{Shorrocks:1978}, cited by \citet{Geweke:Marshall:and:Zarkin:1986:Mobility}: 
\begin{Lemma-InOrder}[Markov Chain Harmonic Mean and the Trace of the Transition Matrix]\label{Lemma:HarmonicMean}
\ 

For a Markov chain with transition matrix $\P$, let $\tau_i$ be the expected duration of state $i$ (the mean length of runs of $i$), and let $E_H(\tau_i)$ be their unweighted harmonic mean.   Let $\lambda_i(\P)$ be the eigenvalues of $\P$.  These are related by the following:
\an{\label{eq:HarmonicMean}
E_H(\tau_i) \eqdef \frac{1}{\displaystyle \frac{1}{n} \sum_{i=1}^n \dspfrac{1}{\tau_i} } 
= \frac{1}{\displaystyle  1 - \frac{1}{n} \sum_{i=1}^n \lambda_i(\P)} \geq 1,
\intertext{or, equivalently}\label{eq:HarmonicMean2}
E(\lambda_i(\P) ) \eqdef \frac{1}{n} \sum_{i=1}^n \lambda_i(\P)   =  1 - \frac{1}{E_H(\tau_i)}  \geq 0.
}
\end{Lemma-InOrder}
\begin{proof}
The expected length of runs of any environmental state $i$ is \citep{Prais:1955:Measuring}
\ab{
\tau_i &\eqdef E(\text{duration of i}) = \sum_{t=0}^\infty t {P}_{ii}^{t-1} (1-{P}_{ii}) \\
&= \sum_{t=0}^\infty (t+1) {P}_{ii}^{t}  - \sum_{t=1}^\infty t {P}_{ii}^{t} 
= \sum_{t=0}^\infty  {P}_{ii}^{t}
= \frac{1}{1-{P}_{ii}}.
}
Since $P_{ii} \geq 0$, then $\tau_i \geq 1$ so $E_H(\tau_i) \geq 1$.  Since the trace of the matrix is
$
\sum_{i=1}^n {P}_{ii} = \sum_{i=1}^n \lambda_i(\P),  
$ we have
\ab{
\sum_{i=1}^n \lambda_i(\P) = \sum_{i=1}^n {P}_{ii} = \sum_{i=1}^n \left(1- \frac{1}{\tau_i}\right)
= n - \sum_{i=1}^n  \frac{1}{\tau_i} \geq 0.
} 
Rearrangements of the terms gives \eqref{eq:HarmonicMean} and \eqref{eq:HarmonicMean2}.
\end{proof}

The main result is now presented, tying together the eigenvalues of the environment transition matrix, $\P$, the effect of dispersal on the population growth rate, and the harmonic mean of environment durations.  The result gives sufficient conditions in terms of the eigenvalues of $\P$ for departures from the reduction phenomenon.

This theorem must sacrifice some generality in the environment transition matrices in order to obtain tractability.  The environmental change process must be a reversible Markov chain, which means it does not exhibit directional cycles (which requires complex eigenvalues in $\P$), which might be thought of as `currents' through the set of states.  This is the symmetrizability constraint that appears in Karlin's Theorem 5.1, and is required for technical reasons described in \emph{Methods}.  It remains an open problem whether the following theorem extends to all ergodic Markov chains.

\begin{Theorem-InOrder}[Eigenvalues, Reduction Phenomenon, and Harmonic Mean of Environment Durations]\label{Theorem:Main}
\ 

Let $\P$ and $\Q \in \Reals^{n,n}$ be transition matrices of reversible ergodic Markov chains that commute with each other.  Let $\tau_i = 1/(1 - P_{ii})$ be the expected length of runs of state $i$ under iteration of $\P$. Let
\an{\label{eq:M}
\M(m) \eqdef \P [(1-m)\I + m\Q],
}
and $\D \neq c \, \I$ be a positive diagonal matrix. 

\enumlist{
\item If all eigenvalues of $\P$ are positive, then
\an{\label{eq:EigPos}
\df{}{m}\rho(\M(m) \D)  < 0,
\intertext{(the reduction phenomenon) and}
E_H(\tau_i ) > 1 + \frac{1}{n-1} \label{EHMore}.
}
\item If all eigenvalues of $\P$ other than the Perron root $1$ are negative, then \label{item:EigNeg}
\an{\label{eq:EigNeg}
\df{}{m}\rho(\M(m) \D)  > 0,
\intertext{(departure from the reduction phenomenon) and}
1 \leq E_H(\tau_i ) < 1 + \frac{1}{n-1} \label{EHLess}.
}
}	
\end{Theorem-InOrder}
The proof is given in \emph{Methods} section \ref{section:proof:Main}.

{\bf Remark}.  One should be careful here not to interpret this result as an implication \emph{from} $E_H(\tau_i )$ \emph{to} $\dfl{\rho(\M(m) \D)}{m}$.  While it would be ideal to derive conditions for the reduction phenomenon from conditions on the durations of the environments, this is not possible here for $n \geq 3$;  rather, both implications derive from the condition on the eigenvalues.  

However, for $n=2$ environments, the implication becomes possible, as seen in this slight generalization of \citet[Theorem B Corollary]{\McD}:
\begin{Corollary-InOrder}\label{Corollary:n=2}
Let $\P$ and $\Q$ be $2\times 2$ irreducible stochastic matrices that commute, and $\D \neq c \, \I$ be a positive diagonal $2\times 2$ matrix.  Then
\ab{
\df{ }{m} \rho(\P [(1-m) \I + m \Q] \D > 0,
}
if and only if  $P_{12} + P_{21} > 1$, or equivalently, $E_H( \tau_1, \tau_2) < 2$.
\end{Corollary-InOrder}
\begin{proof}
In the case of $n=2$, there is only one other eigenvalue, $\lambda_2 = 1 - (P_{12} + P_{21})$.  We have 
\ab{
E_H(\tau_1, \tau_2) = \frac{1}{\displaystyle  1 - \frac{1}{2}[1 + 1 - (P_{12} + P_{21})]}
=  \frac{2}{P_{12} + P_{21}}.
}
So $P_{12} + P_{21} > 1 \iff E_H( \tau_1, \tau_2) < 2 \iff \lambda_2 < 0$, and by Theorem \ref{Theorem:Main}, $\df{\rho(\M \D) }{m} > 0$. 
\end{proof}

The model \eqref{eq:McD2} of \McDA\ is a special case of Theorem \ref{Theorem:Main} and Corollary \ref{Corollary:n=2}.  Note that all irreducible stochastic $2 \times 2$ matrices are transition matrices of ergodic reducible Markov chains:
\begin{Corollary-InOrder}\label{Corollary:SpecialCases}
Theorem \ref{Theorem:Main} includes, as special cases, $\Q = {\P}^t$ for $t \geq 1$, and $\Q = {\P}^\infty = \piv \evt$, where $\piv = \P \piv$ is the stationary distribution of $\P$.
\end{Corollary-InOrder}
\begin{proof}
${\P}$ and $ {\P}^t$ commute, as do ${\P}$ and $\P^\infty = \piv \evt$, since $\P \piv \evt =  \piv \evt = \piv \evt  \P$.  When $\P$ is the transition matrix of a reversible ergodic Markov chain, so too are $\P^t$ and ${\P}^\infty$.
\end{proof}

Theorem \ref{Theorem:Main} is able to give us results only for the extrema of the distribution of eigenvalues of the environment transition matrix, where all non-Perron eigenvalues are either positive or are all negative.  It would be desirable to obtain some results for the interior region where there a mixture of positive and negative eigenvalues.  Short of this, the location of the interior region can at least be made precise in terms of its relation to the space of matrices whose non-Perron eigenvalues are all negative.  

This is done by forming a path from such extreme $\P$ to the opposite extreme, $\I$ --- a homotopy $[0,1] \mapsto \{ \M\} \subset \Reals^{n \times n}$.  For $\M(m)$, the two endpoints of the path are $\M(0, m) =  (1-m)\I + m\Q$, and $\M(1, m) =  \P [(1-m)\I + m\Q]$.  The path can be easily created by a convex combination, $\M(\alpha, m)  = [(1-\alpha) \I + \alpha\P] [(1-m)\I + m\Q]$, parameterized by $\alpha \in [0,1]$.   It is straightforward from Lemma \ref{Lemma:HarmonicMean} that 
\ab{
E_H(\tau_i([1-\alpha] \I + \alpha\P) = \frac{1}{\alpha} E_H(\tau_i(\P)),
}
so as $\alpha$ goes to $0$, the harmonic mean of the environment durations goes to infinity.  Using the homotopy between two extremes, we will now see that, for any $\P$, there is always some $\alpha$ below which the reduction phenomenon holds.

\begin{Corollary-InOrder}[Convex Combination with an Extreme $\P$]\label{Corollary:DistanceFromIdentity}
\ 

Let
\ab{
\M(\alpha, m) \eqdef  [(1-\alpha) \I + \alpha\P] [(1-m)\I + m\Q],
}
where $\P$ and $\Q$ be transition matrices of ergodic reversible Markov chains that commute with each other, and $\alpha, m \in [0,1]$.  Let $\D \neq c \, \I$ be a positive diagonal matrix.

Suppose that $\pf{}{m}\rho(\M(1,m)\D) > 0$ for $m \in (0, 1]$.  Then there exist critical values $\alpha_0, \alpha_1$ with $1/2 \leq \alpha_0 \leq \alpha_1 < 1$, such that for $m \in (0,1]$,
\ab{
\pf{}{m}\rho(\M (\alpha, m) \D)  < 0 \text{\ \ for } \alpha \in [0, \alpha_0) ,
\intertext{and}
\pf{}{m}\rho(\M (\alpha, m) \D)  > 0 \text{\  \ for }  \alpha \in (\alpha_1, 1].
}

\end{Corollary-InOrder}
\begin{proof}
$\M(\alpha,0)=(1-\alpha) \I + \alpha\P$ plays the role of $\P$ in Theorem \ref{Theorem:Main}, so we need to know when $\lambda_i(\M(\alpha,0)) > 0 \ \forall i$.  Let $\lambda_{\min}(\P) \eqdef \min_{i} \lambda_i(\P)$.  Then  
\ab{
\lambda_{\min}&([(1-\alpha) \I + \alpha\P]) = (1-\alpha) + \alpha \lambda_{\min}(\P) > 0
\iff  \\
 \lambda_{\min}&(\P) > (\alpha - 1)/\alpha = 1 - 1/\alpha.
}
By Perron-Frobenius theory, irreducible stochastic $\P$ means $\lambda_{\min}(\P) > -1$.  So $-1 > 1 - 1/\alpha$ (i.e. $\alpha \leq 1/2$) assures $\lambda_i(\M(\alpha,0)) > 0 \ \forall i$.  Thus $\alpha_0$ is no smaller than $1/2$.

Since $\rho(\M (\alpha, m) \D)$ is a continuous function of $\alpha$ and $m$, we know $\pfinline{\rho(\M (\alpha, m) \D)} {m}  > 0$ for $\alpha$ in some neighborhood $(\alpha_1, 1]$. 
\end{proof}
So now we have characterized the interior regions as $[a_0, a_1]$ where the behavior of $\pfinline{\rho(\M(\alpha, m) \D)}{m}$ needs to be characterized.  We do not know, for example, if $\pfinline{\rho}{m}$ keeps the same sign for all $m$ at a given $\alpha$, or whether it can change sign more than once on $[\alpha_0, \alpha_1]$, and so forth.  It remains an open problem.

Next, a particular kind of environmental change process is considered in which there is no causal connection between sequential environments.  In other words, when the environment changes, it has no memory of its previous state.  In genetics this is Kingman's `House of Cards' model of mutation \citep{Kingman:1978,Kingman:1980}).  It is shown for such a memoryless environment that the reduction phenomenon is the only possible outcome.
\begin{Theorem-InOrder}[`House of Cards' Environmental Change]\label{Theorem:HouseOfCards}
\ 

Let $\M(m)$ be defined as in \eqref {eq:M} of Theorem \ref{Theorem:Main}.  Let $\D \neq c\,\I$ be a positive diagonal matrix.  Suppose that when the environment changes, its current state has no influence on its next state.  Suppose further that the expected duration of an environment is $\tau_i = \tau$ for all environments $i$.   

If $\tau = 1$, then 
$
\df{}{m}\rho(\M(m) \D)  = 0. \ \\
$

If $\tau > 1$, then 
$
\df{}{m}\rho(\M(m) \D)  < 0.
$
\end{Theorem-InOrder}
The proof is given in \emph{Methods} section \ref{section:proof:HouseOfCards}.   Note that in this case, $\tau >1$ becomes a sufficient condition for selection for reduced dispersal, which it is not in the general case in Theorem \ref{Theorem:Main}.

\subsection{The Conditional Dispersal Model}
The result of \citet{\McD} that drew particular attention was their finding that, under a broad range of circumstances, it is better for the organism to ignore cues about the environment and instead follow philopatry.  The  general form for their cue model is a modification of \eqref{eq:McDgeneral}:
\ab{
\M \D &= \P [ (\I - \C) + \piv \evt \C] \D  ,
}
where $\C$ is a diagonal matrix of the conditional dispersal probabilities, $C_{i}$, that an individual disperses given it is in environment $i$.  

A change in an organism's response to cues about its environment is manifest as changes to $C_j$, hence the object of interest is the change in asymptotic growth as the conditional dispersal rate is changed:
\ab{
\pf{}{C_j}  \rho(\P [ (\I - \C) + \piv \evt \C] \D).
}
The next result shows that an organism should increase its dispersal from any environment where its destinations correlate better with left Perron vector $\uv(\M\D)$ than does staying put, and that in general there is \emph{always} at least one such environment.

\begin{Theorem}[Conditional Dispersal]\label{Theorem:Cues}
\  \\
Let $\M \eqdef \P [ (\I - \C) + \piv \evt \C]$, where $\P$ be an irreducible stochastic matrix, $\C$ and $\D$ are positive diagonal matrices, with $C_i \in (0, 1)$, and $\piv = \P \piv$.  Refer to the left and right Perron vectors as $\uv \equiv \uv(\M\D)$ and $\vv \equiv \vv(\M\D)$.  Then:
\enumlist{
\item  The derivative of the spectral radius with respect to each $C_\kappa$ is
\an{
\pf{}{C_\kappa}&  \rho(\P [ (\I - \C) + \piv \evt \C] \D) \notag \\
&=  D_\kappa \,  v_\kappa \sum_{i=1}^n u_i (  \pi_i -   P_{i\kappa} ) \label{eq:SumuPiP}\\&
=  D_\kappa  v_\kappa n \left[  \Cov(u_i, \pi_i ) -  \Cov(u_i, P_{i\kappa} ) \right] . \notag
} 
\item \label{item:Kappas} There is always at least one $\kappa$ for which
\ab{
\pf{}{C_\kappa}  \rho(\P [ (\I - \C) + \piv \evt \C] \D) > 0,
\intertext{and at least one $\kappa$ for which}
\pf{}{C_\kappa}  \rho(\P [ (\I - \C) + \piv \evt \C] \D) < 0,
}
unless $\D = c \, \I$ or $\P = \piv \evt$.
\item \label{item:Gradient} The gradient of the spectral radius with respect to $\C$ is:
\ab{
\Gradient_\C \ \rho(\M\D) &\eqdef \matrx{\pf{\rho(\M\D)}{C_\kappa}}_n^{\kappa=1}
=  \uv\tr (\piv \vv\tr  -  \P \D_\vv ) \D.
}
\item  \label{eq:NeutralContour} There is always a subspace, $\Nc$, of perturbations of $\C$ that are neutral for $\rho(\M\D)$.  $\Nc = \{\xiv \suchthat \Gradient_\C \, \rho(\M\D)\  \xiv = 0 \}$ is an $n-1$ dimensional linear subspace.  Its basis includes strictly positive $\zetav = \D_v^{-1} \D^{-1} \piv$, i.e. 
\an{\label{eq:NeutralLine}
\zeta_\kappa =  \dspfrac{\pi_\kappa}{D_\kappa v_\kappa}.
}
}	
\end{Theorem}
The proof is given in \emph{Methods} section \ref{section:proof:Cues}.

Theorem \ref{Theorem:Cues} shows that an organism can increase its asymptotic growth rate by dispersing more from any environment $\kappa$ for which $\Cov(u_i(\M\D), \pi_i ) > \Cov(u_i(\M\D), P_{i\kappa} )$, which means that its environment if it disperses (distributed as $\piv$), correlates better with $\uv(\M\D)$ than its environment if it stays put (distributed as $[\P]_\kappa$).  

Furthermore, there is always at least one such environment where \emph{increased} dispersal is advantageous, and at least one other environment where \emph{decreased} dispersal is advantageous, unless dispersal is neutral.  Dispersal is neutral  when growth rates are the same in all environments ($\D= c \, \I$), or the present environment has no influence on the next environment ($\P = \piv \evt$).

We see that Theorem \ref{Theorem:Cues} provides another situation that departs from the reduction phenomenon.  Conditional dispersal is analogous to directed mutation \citep{Cairns:Overbaugh:and:Miller:1988,Hall:1990,Lenski:and:Mittler:1993}.  I would not go so far as to say it exemplifies the principle of partial control, which was conceived for the situation of multiple \emph{undirected} transformation processes, such as recombination in the presence of mutation.  It is possible, however, to view conditional dispersal as control over only a part of the set of dispersal probabilities.

The existence of this departure from reduction holds for any environment transition matrix $\P \neq \piv \evt$.  The environment may even be constant, $\P = \I$, in which case as long as $\D \neq c\,\I$ so $\uv \neq \ev$, then $\pf{}{C_\kappa} \rho( [ (\I - \C) + \piv \evt \C] \D) > 0$ for every $\kappa$ where $u_\kappa$ is below the $\piv$-weighted average $\sum_{i=1}^n u_i   \pi_i $ (see \eqref{eq:SumuPiP}).

\begin{figure}[top]
\includegraphics[width=3.6in]{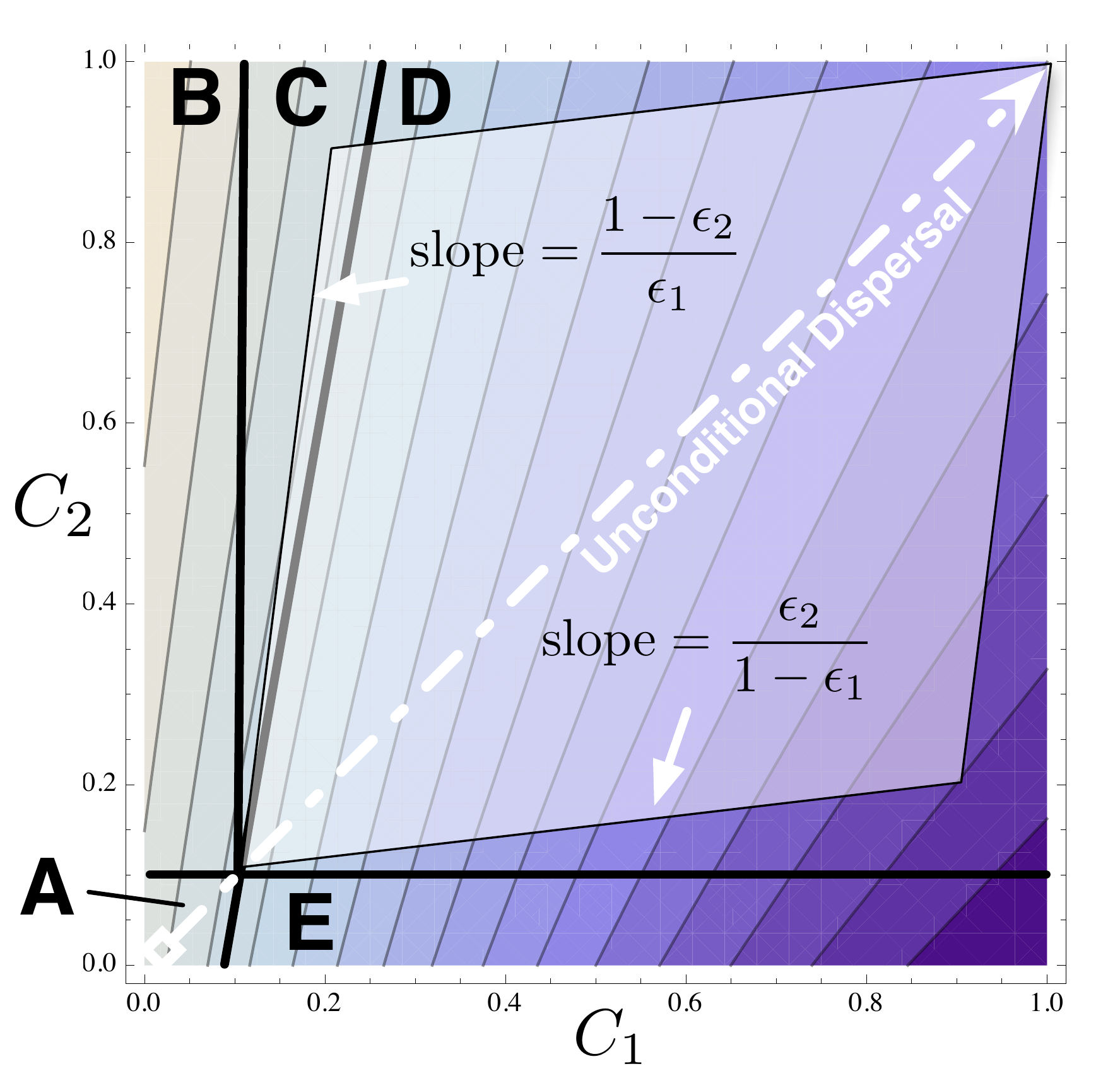}
\caption{\small\label{fig:ConditionalDispersal} Gradient of the asymptotic growth rate, $\rho(\M\D)$, over $(C_1, C_2) \in [0,1]^2$.  Lighter means higher $\rho(\M\D)$.  Model parameters:  $(D_1, D_2) = (1.0, 0.5)$, $(p_{12}, p_{21}) = (0.204, 0.107)$.  Diagonal dashed line $C_1\!=\!C_2$ corresponds to variation in unconditional dispersal rates.  Perturbations are away from $(C_1, C_2) = (0.1, 0.1)$.  Regions A+B+C increase $\rho(\M\D)$.  Regions D+E decrease $\rho(\M\D)$.  \McDA\ constrain variation to fall within the parallelogram, with slope $\epsilon/(1-\epsilon)$ for the bottom, and $(1-\epsilon)/\epsilon$ for the side, where $\epsilon$ is the error rate for environmental cues;  the error rate must be small enough for the parallelogram to enter region C for conditional dispersal to evolve.  But the \emph{unconstrained} ESS here is $(C_1, C_2) = (0, 1)$, and mutants anywhere in regions B+C increase conditional dispersal rate $C_2$ and are advantageous.
}	
\end{figure}
Philopatry is clearly not the evolutionarily stable state (ESS) when there is environmental change, since there is always at least one environment where conditional dispersal is advantageous.  To understand how \McDA\ conclude that philopatry can be an ESS, it is helpful to look at the entire `adaptive landscape' of $\rho(\M\D)$ as a function of conditional dispersal rates $(C_1, C_2)$.  

\McDA\ posit a species that can vary its probabilities, $(p_1, p_2)$, of dispersing in response to a binary cue, 
where environment $i$ produces the wrong cue with probability $\epsilon_i$.  The conditional dispersal rates are thus
\an{\label{eq:CueModel}
\Bmatr{C_1\\C_2} = \Bmatr{1-\epsilon_1 & \epsilon_1\\ \epsilon_2 & 1- \epsilon_2}\Bmatr{p_1\\p_2} + \Bmatr{C_{\min}\\C_{\min}},
}
Assuming that the species can vary $(p_1, p_2)$ over the range $[0,1]\times[0,1]$, the variation in $(C_1, C_2)$ falls within the parallelogram in Figure \ref{fig:ConditionalDispersal}. 

The example depicted in Figure \ref{fig:ConditionalDispersal} has a moderate rate of environmental change and differential growth between two environments.  Darker means smaller $\rho(\M\D)$. Variation along the white diagonal line represents unconditional dispersal, and the gradient exhibits the reduction phenomenon.  The contour lines of constant $\rho(\M\D)$ are shown by \eqref {eq:NeutralLine} to have slope 
\ab{
\left(\dspfrac{\pi_2}{D_2 v_2}\right) / \left(\dspfrac{\pi_1}{D_1 v_1}\right)
= {\dspfrac{\pi_2 D_1 v_1}{\pi_1  D_2 v_2}}.
}
The dispersal rates of the resident population are $(C_1, C_2) = (0.1, 0.1)$, the minimal dispersal attainable.  The labeled regions A, B, C, D, and E demarcate the behavior where $(C_1, C_2)$ departs from $(0.1, 0.1)$.  Any mutant that falls within regions A, B, or C is advantageous.  Regions B and C comprise increases in conditional dispersal from environment $C_2$.  The slopes of the sides of the parallelogram derive from the cue error rate, $\epsilon$.

Advantageous mutants arise only from in the intersection of the parallelogram and region C.  For the intersection to be non-empty, $\epsilon$ must be small enough.  When the error rate is so high that the parallelogram is contained entirely in region D, then the ESS is the lower left corner, the minimal value of dispersal.

We can see that the ESS is very sensitive to the error rate, however.  With a slight decrease in $\epsilon$, the ESS can shift from the lower left corner to the upper left corner of the parallelogram, which describes \McDA's result.  

Therefore, in this adaptive landscape, it becomes very important how well the genetic and developmental system fills out the parallelogram with heritable variation.  If the variation does not fully fill it out, at least two novel outcomes become possible.  A slightly convex distribution overlapping region C would yield an intermediate level of dispersal as the fittest that occur.  A slightly concave distribution would result in a bimodal distribution of the fittest phenotypes.  This opens up potential for polymorphisms, disruptive selection, history dependence, or evolutionary volatility in the phenotype.  

Moreover, the parallelogram is based on a particular model \eqref{eq:CueModel} of how organisms disperse.  There is no categorical exclusion of genetic variation from accessing any point in the square $(C_1, C_2) \in [0, 1]^2$.  Rather, it depends on the details of the organism, its capabilities, and its genotype-phenotype map.  Any time an evolutionary outcome is sensitive to such details, one is bound to find interesting phenomena in the natural history.

\section{Methods} \label{section:Methods}
{
The lengthier proofs for Theorems \ref{Theorem:Main}, \ref{Theorem:HouseOfCards}, and \ref{Theorem:Cues} are now provided, prefaced by preparatory results, Lemma \ref{Lemma:CanonicalForm} and Theorem \ref{Theorem:SpectralRadius}.  It is here that we encounter the tractability afforded by using the transition matrices of reversible ergodic Markov chains, which is the key technique adopted from \citet[Theorem 4.1]{Friedland:and:Karlin:1975} and Karlin's Theorem 5.1.  

It should be noted that Karlin's Theorem 5.2 is not usable here, because of the presence of matrix $\P$ in $\M(m) = \P[(1-m)\I + m \Q]$.  This is why it has been an open problem \citep{Altenberg:2004:Open}.  An initial inroad on this open problem was obtained through application of elements from Karlin's Theorem 5.1 to the analysis of multivariate, multiple locus mutation rate evolution in \citet{Altenberg:2011:Mutation}.  The application of these techniques is further extended here.  Theorem 2 in \citet{Altenberg:2011:Mutation} --- a multivariate reduction principle for multiple loci in mutation-selection balance --- is in fact a special case of \eqref{eq:EigPos} in Theorem \ref{Theorem:Main} given here.  

In the proofs to follow,  \eqref{eq:S}, \eqref{eq:xVariational}, \eqref{eq:xhat} derive from Karlin's Theorem 5.1 proof.  Other steps, including the use of a canonical form for symmetrizable $\M(m)$ \eqref {eq:CanonicalA}, \eqref{eq:K}, \eqref{eq:B}, and \eqref{eq:drho-y2} are drawn from the analysis in \citet{Altenberg:2011:Mutation}.  Most of the remaining steps arise naturally from the problem, and may prove useful in other contexts.

Theorem \ref{Theorem:Main} first requires a characterization of the spectral radius of $\M(m) \D$, which relies on the canonical form for $\M(m)$ that exists when it is constrained to be symmetrizable.
%
\begin{Lemma-InOrder}[Canonical Form for Symmetrizable $\M(m)$]\label{Lemma:CanonicalForm}
\ 

Let $\P$ and $\Q \in \Reals^{n,n}$ be transition matrices of ergodic reversible Markov chains that commute with each other.  Let
\ab{
\M(m) \eqdef \P [(1-m)\I + m\Q].
}
Then $\P$, $\Q$, and $\M$ can be decomposed as
\an{
\P &= \D_{\piv}^{1/2}\K \Lam_P \K\tr \D_{\piv}^{-1/2}, \label{eq:CanonicalP}\\
\Q &= \D_{\piv}^{1/2} \K \Lam_Q \K\tr \D_{\piv}^{-1/2}, \label{eq:CanonicalQ}
\intertext{and}
\M(m) &=  \D_{\piv}^{1/2} \K \Lam_P[(1-m) \I + m \Lam_Q] \K\tr \D_{\piv}^{-1/2}, \label{eq:CanonicalM}
}
where 
$\P \piv = \Q \piv = \piv$, with $\evt \piv = 1$, $\K$ is an orthogonal matrix, and $\Lam_P$ and $\Lam_Q$ are diagonal matrices of the eigenvalues of $\P$ and $\Q$, respectively.

Furthermore, the first column of $\K$ is $[\K]_1 = \piv^{1/2}$ (element-wise square root).
\end{Lemma-InOrder}

\begin{proof}
The transition matrices of ergodic reversible Markov chains, $\M$, can be represented in a canonical way (\citealt[p. 33]{Keilson:1979}; \citealt[p. 296]{Ababneh:Jermiin:and:Robinson:2006}; \citealt[Lemmas 1 and 2]{Altenberg:2011:Mutation}) as 
\an{\label{eq:CanonicalA}
\M = \B \K \Lam \K\tr \B^{-1},
}
where $\B$ is a positive diagonal matrix, unique up to scaling, and $\K$ is an orthogonal matrix, i.e. $\K \K\tr =\K\tr \K = \I$.  

Any such $\M$ is clearly diagonalizable since $\Lam$ is a diagonal matrix.  Diagonalizable $\P$ and $\Q$ commute by hypothesis, so they can be simultaneously diagonalized \citep[Theorem 1.3.19, p. 52]{Horn:and:Johnson:1985}, which means there exists invertible $\X$ such that
\ab{
\P = \X \Lam_P \X^{-1} \text{\ and \ } \Q = \X \Lam_Q \X^{-1}.
}
Hence
$
\M(m) = \X \Lam_P[ (1-m) \I + m \Lam_Q] \X^{-1}.
$
Combining these two forms, the common matrix $\X$ can be represented as $\X = \B \K$.

Next, it is shown that
\an{ 
[\K]_1 &= \piv^{1/2}  \label{eq:K}
\intertext{ and } 
\B &= c \ \D_{\piv}^{1/2}, \ c > 0, \label{eq:B}
}
satisfy $\evt \P = \evt$ and $\P \piv = \piv$.
Since $\K$ is orthogonal, $[\K]_1 = \piv^{1/2}$ if and only if ${(\piv^{1/2})}\tr \K = \ev_1\tr$, in which case, recalling that $\lambda_{P1} = 1$, substitution gives 
\ab{
\evt \P &=  \evt \D_{\piv}^{1/2} \K \Lam_P \K\tr \D_{\piv}^{-1/2}
= \ev_1\tr \Lam_P \K\tr \D_{\piv}^{-1/2}  \notag \\
&= \ev_1\tr  \K\tr \D_{\piv}^{-1/2} 
= [\K]^1 \D_{\piv}^{-1/2} = \evt , 
\intertext{and} 
\P \piv &= \D_{\piv}^{1/2} \K \Lam_P \K\tr  \D_{\piv}^{-1/2} \piv 
=  \D_{\piv}^{1/2} \K \Lam_P \K\tr \piv^{1/2} \notag\\
&=  \D_{\piv}^{1/2} \K \Lam_P \, \ev_1  \notag 
= \D_{\piv}^{1/2} [\K]_1 \notag
= \piv. \notag
}
Substitution of $\B = c \ \D_{\piv}^{1/2}$ in the form \eqref{eq:CanonicalA} produces \eqref{eq:CanonicalP}, \eqref{eq:CanonicalQ}, and \eqref{eq:CanonicalM}.

{\bf Remark}.  For any family of commuting symmetrizable stochastic matrices,   $\K$ and $\B$ (up to scaling) are uniquely determined.  Therefore, the only variation possible for the family is in $\lambda_i$, $i = 2, \ldots, n$, which means there are at most $n-1$ degrees of freedom of variation in the family.
\end{proof}

\begin{Theorem-InOrder}[The Spectral Radius]\label{Theorem:SpectralRadius}
\ \\
Let $\P$ and $\Q \in \Reals^{n,n}$ be transition matrices of ergodic reversible Markov chains that commute with each other, let $\piv$ be their common right Perron vector, and let $\{ \lambda_{Pi} \}$ and $\{ \lambda_{Qi} \}$ be their eigenvalues.  Let
\ab{
\M(m) \eqdef \P [(1-m)\I + m\Q].
}
Let $\D$ be a positive diagonal matrix.   Set $\vv \equiv \vv(\M(m) \D)$ and $\uv \equiv \uv(\M(m) \D)$.

Then
\an{\label{eq:rho-y2}
\rho(\M(m) \D) &=  \sum_{i=1}^n \lambda_{Pi} [(1-m) + m \lambda_{Qi} ] y_i^2 ,
}
where
\ab{
\y &=  (\vv\tr \ \D_\piv^{-1} \D \vv )^{-1/2} \ \K\tr \D_{\piv}^{-1/2} \D \, \vv,
}
and $\K$ is from the canonical form in Lemma \ref{Lemma:CanonicalForm}.

The left and right Perron vectors of $\M(m)\D$ are related by
\ab{
\uv &= \frac{1}{(\vv \tr  \D_\piv^{-1}  \D \vv)}  \D_\piv^{-1}  \D \ \vv.
}
\end{Theorem-InOrder}
\begin{proof} 
Canonical form \eqref{eq:CanonicalM} is used to produce a symmetric matrix similar to $\M(m) \D$, which allows use of the Rayleigh-Ritz formula for the spectral radius.    The expression simplifies to a sum of terms involving the eigenvalues of the stochastic matrices $\P$ and $\Q$.  

For brevity let $\Phim \eqdef \K \Lam_P [(1-m) \I  + m \Lam_Q] \K\tr$, so $\M(m) = \B \Phim \B^{-1}$.  Multiplication by $\B$, $\D^{1/2}$, and their inverses (where the positive diagonal $\D$ ensures the existence of $\D^{1/2}$ and $\D^{-1/2}$) gives the identities:
\ab{
\rho(\M(m) \D) &= \rho(  \B \Phim \B^{-1} \D ) = \rho(  \Phim \B^{-1} \D \B)  \\
&= \rho( \Phim \D )  = \rho(\D^{1/2} \Phim \D^{1/2} ) = \rho(\S),
}
where
\an{\label{eq:S}
\S &\eqdef \D^{1/2} \Phim \D^{1/2} \notag \\
&=  \D^{1/2} \K \Lam_P[(1-m) \I + m \Lam_Q] \K\tr \D^{1/2}.
}
Since $\S$ is symmetric, we may apply the Rayleigh-Ritz variational formula for the spectral radius \citep[Theorem 4.2.2, p. 176]{Horn:and:Johnson:1985}:
\an{\label{eq:RayleighRitz}
\rho(\A) = \max_{\x\tr \x = 1} \x\tr \A \x.
}
This yields
\an{\label{eq:xVariational}
\rho&(\M(m) \D) = \notag\\
&\max_{\x\tr \x = 1}  \x\tr \D^{1/2}  \K \Lam_P [(1-m) \I + m \Lam_Q] \K\tr \D^{1/2} \x.
}
Since $\M$ is irreducible and $\D$ a positive diagonal matrix, $\M\D$ is irreducible, so by Perron-Frobenius theory there is a unique eigenvector $\xvh > 0$  that yields the maximum in \eqref{eq:xVariational}, allowing us to write
\an{\label{eq:xhat}
\rho&(\M(m) \D) \notag \\ &
=  \xvh\tr \D^{1/2}  \K \Lam_P [(1-m) \I + m \Lam_Q] \K\tr \D^{1/2} \xvh.
}

Define 
\an{\label{eq:ydef}
\y \eqdef \K\tr  \D^{1/2}  \xvh.
}
Substitution of \eqref {eq:ydef} into \eqref{eq:xhat} yields \eqref {eq:rho-y2}:
\ab{
\rho(\M(m) \D) &=  \xvh\tr \D^{1/2}  \K \Lam_P [(1-m) \I + m \Lam_Q] \K\tr \D^{1/2} \xvh \notag \\
&= \y\tr  \Lam_P [(1-m) \I + m \Lam_Q] \y \notag \\
&= \sum_{i=1}^n \lambda_{Pi} [(1-m) + m \lambda_{Qi} ] y_i^2 . 
}
Next, $\y$ will be solved in terms of $\vv$ by solving for $\xvh$, using the following two facts.  For brevity, define $\Lam_{(m)} \eqdef  \Lam_P [(1-m) \I + m \Lam_Q]$, and write $\M \equiv \M(m) =  \B \K  \Lam_{(m)} \K\tr \B^{-1}$:  
\an{
1.& \ \rho(\M\D) \ \vv = \M \D \vv
=  \B \K  \Lam_{(m)} \K\tr \B^{-1} \D \vv;  \label{fact:v} \\
2.& \ \rho(\M \D) \ \xvh =  \D^{1/2}  \K  \Lam_{(m)} \K\tr \D^{1/2} \xvh. \label{fact:x}
}

Multiplication on the left by $\B \D^{-1/2}$ in \eqref{fact:x}, and substitution of \eqref{fact:v} reveals the right Perron vector of $\M\D$:
\an{
\rho(\M \D) \ &( \B \D^{-1/2}) \xvh  
= \ ( \B \D^{-1/2}) \D^{1/2}  \K \Lam_{(m)}  \K\tr  \D^{1/2} \xvh \notag \\
= &\  \B  \K \Lam_{(m)} 
 \K\tr ( \B^{-1} \D \B \D^{-1}) \D^{1/2} \xvh) \notag \\
= &\ ( \B  \K \Lam_{(m)} \K\tr \B^{-1} \D ) ( \B \D^{-1/2} \xvh) \notag \\
= &\ \M \D ( \B \D^{-1/2} \xvh) , \label{eq:eigvecDerivation}
}
which shows that $\B \D^{-1/2} \xvh$ is the right Perron vector of $\M \D$, unique up to scaling, i.e.
\ab{
\vv = \B \D^{-1/2} \xvh 
=  \ch \, \D_\piv^{1/2} \D^{-1/2} \xvh ,
}
for some $\ch$ to be solved.  This almost finishes the solution of $\xvh$, giving
\an{\label{eq:x=DDv}
\xvh =  \frac{1}{\ch} \D_\piv^{-1/2}  \D^{1/2}  \ \vv .
}
The constraint $\xvh\tr \xvh = 1$ gives
\ab{
1 &= \xvh\tr \xvh =  \frac{1}{\ch^2} (\vv\tr \D_\piv^{-1}  \D  \ \vv),
}
so
\an{\label{eq:ch}
\ch &= (\vv\tr \ \D_\piv^{-1} \D \vv )^{1/2}.
}
Substitution for $\xvh$ now produces the expression in the theorem,
\an{\label{eq:yvec}
\y &\eqdef \K\tr  \D^{1/2}  \xvh
= \K\tr  \D^{1/2}  \frac{1}{\ch} \D_\piv^{-1/2}  \D^{1/2}  \ \vv \notag \\&
=  (\vv\tr \ \D_\piv^{-1} \D \vv )^{-1/2} \ \K\tr  \D_\piv^{-1/2}  \D  \ \vv .
}

By the same method as \eqref{eq:eigvecDerivation}, $\uv(\M(m)\D)$ is derived, using multiplication on the right by $\D^{-1/2}\B $ to reveal the left Perron vector of $\M\D$:
\ab{
\rho(\M \D) \ & \xvh\tr   ( \D^{1/2} \B^{-1} )
=\xvh\tr \D^{1/2}  \K \Lam_{(m)}  \K\tr  \D^{1/2} ( \D^{1/2} \B^{-1} ) \\
& = \xvh\tr (\D^{1/2} \B^{-1}) \M \D = c^* \, \uv(\M\D)\tr
}
for some $c^* > 0$.  From $\B=c \, \D_\piv^{1/2}$ \eqref{eq:B}, we get $(1/c^*) \xvh\tr \D^{1/2} \D_\piv^{-1/2} =  \uv(\M\D)\tr$. Substituting \eqref{eq:x=DDv} and noting $\uv\tr \vv=1$, we see the simple relationship to $\vv(\M(m)\D)$:
\ab{
\uv(\M(m)\D) &= \frac{1}{(\vv \tr  \D_\piv^{-1}  \D \vv)}  \D_\piv^{-1}  \D \ \vv(\M(m) \D).
}
\end{proof} 

One additional property that stems from the symmetrizability of $\M(m)$ in \eqref{eq:M} is that $\rho(\M(m) \D)$ is convex in $m$.  

\begin{Theorem-InOrder}[Convexity of $\rho(\M(m)\D)$ in $m$]\label{Theorem:Convexity}
\ 

Let $\P$ and $\Q \in \Reals^{n,n}$ be transition matrices of ergodic reversible Markov chains that commute with each other.  Let $\D$ be a positive diagonal matrix and 
\ab{
\M(m) \eqdef \P [(1-m)\I + m\Q].
}

Then $\rho(\M(m) \D)$ is convex in $m$.
\end{Theorem-InOrder}
\begin{proof}
This follows the same lines as in \citet[Theorem F.1, p. 199]{Karlin:1982}.  $\rho(\M(m) \D)= \rho(\S)$ in \eqref{eq:S}, and $\S = (1-m) \A + m \B$, where $\A = \D^{1/2} \K \Lam_P \K\tr \D^{1/2}$ and $\B = \D^{1/2} \K \Lam_P\Lam_Q \K\tr \D^{1/2}$.  The convexity of $\rho( (1-m) \A + m \B)$ is established by Lemma \ref {Lemma:Convexity}, to follow.
\end{proof}

\begin{Lemma-InOrder}[Convexity of the Spectral Radius]\label{Lemma:Convexity}
Let $\A$ and $\B$ be two symmetric matrices with unique eigenvectors $\xvh_A$ and $\xvh_B$ associated with their largest eigenvalue, normalized so ${\xvh_A}\tr \xvh_A = {\xvh_B}\tr\xvh_B = 1$.  

Then $\rho((1-m) \A + m \B)$ is convex in $m$, and strictly convex if $\xvh_A \neq {\xvh_B}$
\end{Lemma-InOrder}
\begin{proof}
By hypothesis $\xvh_A$ uniquely yields the maximum in \eqref{eq:RayleighRitz}, and likewise $\xvh_B$ for $\B$, and $\xvh_h$ for $(1-m) \A + m \B$.  Therefore,
\ab{
\rho((&1-m)  \A + m \B) =\ {\xvh_m}\tr ((1-m) \A + m \B) \xvh_m \\
=&\ (1-m) \, {\xvh_m}\tr \A\xvh_m  + m \, x_m\tr \B \xvh_m \\
 \leq&\ (1-m) \, {\xvh_A}\tr \A \xvh_A  + m \,{\xvh_B}\tr \B \xvh_B
= (1-m) \rho(\A) + m \rho(\B).
}
Equality requires $\xvh_A = \xvh_h = \xvh_B$, because otherwise, ${\xvh_A}\tr \A\xvh_A > {\xvh_h}\tr \A \xvh_h$, or ${\xvh_B}\tr \B\xvh_B > \ {\xvh_h}\tr \B \xvh_h$, either of which produces strict inequality.
\end{proof}

\subsection{Proof of Theorem \ref{Theorem:Main}}\label{section:proof:Main}
Theorem \ref{Theorem:SpectralRadius} is now applied to the derivative of the spectral radius.
The general relation is 
\an{\label{eq:Derivative}
\pf{\rho(\A)}{m} = \uv(\A)\tr \pf{\A}{m} \vv(\A)
}
\citep[Sec. 9.1.1]{Caswell:2000}.

This is derived for the specific case here by differentiating $\S \xvh = \rho(\M\D) \xvh$ (recall $\S$ from \eqref{eq:S}), and then multiplying on the left by $\xvh\tr$.  Set $\rho \equiv \rho(\M\D)$.
\ab{
&\xvh\tr \df{( \S \xvh)}{m} = \xvh\tr \left(\df{\S} {m} \xvh +  \S \df{\xvh} {m}\right) 
= \xvh\tr \df{ \S}{m} \xvh +  \rho \   \xvh\tr \df{\xvh}{m} \\
&= \xvh\tr  \df{}{m}( \rho \xvh ) 
=  \xvh\tr \left( \df{\rho}{m} \ \xvh + \rho \   \df{\xvh}{m} \right) 
= \df{\rho}{m} + \rho \ \xvh\tr   \df{\xvh}{m}.
}
Subtraction of $ \rho(\M\D) \ \xvh\tr   \df{\xvh}{m}$ from both sides and substituting with  \eqref{eq:xhat} leaves:
\ab{
\df{\rho(\M\D)}{m} &= \xvh\tr \df{ \S}{m} \xvh  \\
 =\xvh\tr &   \df{}{m} \left[\D^{1/2}  \K \Lam_P[(1-m) \I + m \Lam_Q] \K\tr \D^{1/2} \right] \xvh \\
 =  \xvh\tr  & \D^{1/2}  \K \Lam_P[\Lam_Q -  \I] \K\tr \D^{1/2} \xvh.
}
Substitution with $\y \eqdef \K\tr  \D^{1/2} \xvh$ yields the derivative of \eqref {eq:rho-y2}:
\an{\label{eq:drho-y2}
\df{\rho(\M \D) }{m} 
= \y\tr  \Lam_P ( \Lam_Q - \I ) \y 
= \sum_{i=1}^n \lambda_{Pi} (\lambda_{Qi} - 1) y_i^2.
}
{\bf Remark}.  Were $\P$ and $\Q$ not symmetrizable, but only diagonalizable and commuting, the analysis would arrive at an expression similar to \eqref{eq:drho-y2} except that the nonnegative $y_i^2$ terms would be replaced by products whose signs we do not know, preventing further evaluation.

We know several things about the terms in the sum in \eqref{eq:drho-y2}:
\enumlist{
\item Since $\P$ and $\Q$ are stochastic matrices, their Perron roots are 1, which here are labelled as $\lambda_{P1} = \lambda_{Q1} = 1$.
\item $\lambda_{Q1} - 1 = 0$.  Thus  the first term of the sum is zero.  \label{item:lambda0}
\item $\lambda_{Qi} - 1 < 0$, for $i \in \{2, \ldots, n\}$, hence $(\lambda_{Qi} - 1) y_i^2 \leq 0$.  

Since $\P$ and $\Q$ are symmetrizable, $\lambda_{Pi}, \lambda_{Qi} \in \Reals$.  Since $\P$ and $\Q$ are irreducible, by Perron-Frobenius theory \citet [Theorems 1.1, 1.5]{Seneta:2006}, eigenvalue 1 has multiplicity 1, and $| \lambda_{Qi} | \leq 1$, which together imply $\lambda_{Qi}  < 1$ for $i \in \{2, \ldots, n\}$. \label{item:lambdaneg}
\item \label{item:y not 0}$y_i \neq 0$ for at least one $i \in \{2, \ldots, n\}$, whenever $\D \neq c \, \I$ for any $c > 0$.  This fact will take a bit of work to show:   \label{item:ypos}
Suppose to the contrary that $y_i = 0$ for all $i \in \{2, \ldots, n\}$.  That means $\y = y_1 \, \ev_1$.  Using $\ch$ from \eqref{eq:ch}, \eqref{eq:yvec} becomes
\ab{
y_1 \, \ev_1 =  \ch^{-1} \ \K\tr  \D_\piv^{-1/2}  \D  \ \vv.
}
Multiplication on the left with $\D_\piv^{1/2} \K$, and substitution with $[\K]_1 = \piv^{1/2}$ \eqref{eq:K} yields
\ab{
y_1 \, \D_\piv^{1/2} \K \ev_1 &= y_1 \, \D_\piv^{1/2} [\K]_1 =   y_1 \, \piv = \ch^{-1}  \D  \ \vv .
}
Multiplication of $y_1 \ch \, \piv =   \D \vv$ by $\M$ gives
\ab{
 \M \piv \, y_1 \ch = y_1 \ch  \piv = \D \vv  =\M \D \vv =  \rho(\M\D) \, \vv.
}
Hence,  $\D \vv = \rho(\M\D) \, \vv$, implying $\D = \rho(\M\D) \, \I$, contrary to hypothesis.  Therefore, $\D \neq c \, \I$ for any $c > 0$ implies that $y_i \neq 0$ for at least one $i \in \{2, \ldots, n\}$. 
}

Points \ref{item:lambdaneg}., and \ref{item:ypos}.\ above together imply that $(\lambda_{Qi} - 1) y_i^2 < 0$ for at least one $i \in \{2, \ldots, n\}$.  Inclusion of point \ref{item:lambda0}.\ immediately implies for \eqref{eq:drho-y2} that:
\enumlist{
\item If $\lambda_{Pi} > 0$ for all $i$, then $\df{ }{m}\rho(\M \D) < 0$.\\
\item If $\lambda_{Pi} < 0$ for $i = 2, \ldots, n$, then $\df{}{m}\rho(\M \D)  > 0$.\\
\item Otherwise: there may be positive, negative, or zero terms $\lambda_{Pi} (\lambda_{Qi} - 1) y_i^2$, so the sign of $\df{ }{m}  \rho(\M \D)$ depends on the particular values of the terms, of which we know little at this point.
}
{\bf Remark}.  Condition $\lambda_{Pi} > 0$ for all $i$ is equivalent to $\P$ being symmetrizable to a positive definite matrix, which is the hypothesized condition in Karlin's Theorem 5.1.  The condition $\lambda_{Pi} < 0$ for $i \in \{ 2, \ldots, n\}$ in case \ref{item:EigNeg}.\ happens to be the same as the well-known condition on the fitness matrix for a stable multiple-allele polymorphism \citep{Kingman:1961:Mathematical}.  Here this appears to be coincidence, rather than a clue to some deeper result.  However, that condition is central to the `viability-analogous' modifier polymorphisms, where the matrix $\left[1-m_{ij}\right]_{i,j=1}^n$ (from diploid modifier genotypes $i | j$), must have all negative eigenvalues except the Perron root to assure stability of the modifier polymorphism \citep{Feldman:and:Liberman:1986,Liberman:and:Feldman:1986:Recombination,Liberman:and:Feldman:1986:Mutation,Liberman:and:Feldman:1989}, supporting the analogy between viability coefficients and modifier values $1-m_{ij}$.

\emph{The Harmonic Mean}.  The following inequalities are equivalent:
\an{
E_H(\tau_i) & \eqdef \frac{1}{\displaystyle \frac{1}{n} \sum_{i=1}^n \dspfrac{1}{\tau_i} } < 1 + \frac{1}{n-1} =\frac{n}{  n -  1} \label{eq:EHLessV2} \\ 
\iff  & n -  1 < \sum_{i=1}^n  \frac{1}{\tau_i} = n -   \sum_{i=1}^n {P}_{ii} \notag\\
\iff  &1  >   \sum_{i=1}^n {P}_{ii} = \sum_{i=1}^n \lambda_i(\P)  = 1 +  \sum_{i=2}^n \lambda_i(\P) \notag\\
\iff  & 0  >   \sum_{i=2}^n \lambda_i(\P) \label{eq:EHLambda}.
} 
Hence, $\lambda_i(\P) < 0$ for $i = 2, \ldots, n$ implies \eqref{eq:EHLambda}, or equivalently, \eqref{eq:EHLessV2}.  Conversely, $\lambda_i(\P) > 0$ for $i = 2, \ldots, n$ implies $\sum_{i=2}^n \lambda_i(\P) > 0$, hence \eqref{EHMore}. \qed

\subsection{Proof of Theorem \ref{Theorem:HouseOfCards}, `House of Cards' Environmental Change}\label{section:proof:HouseOfCards}
By hypothesis, the probability that environment remains unchanged in one generation is $\sigma = 1 - 1 / \tau$.  If the current environment has no influence on which environment comes next (Kingman's `House of Cards' model, \citealt{Kingman:1978}, \citeyear{Kingman:1980})
then
\ab{
P_{ij} = (1- \sigma) \pi_i + \sigma \delta_{ij}
\intertext{where $\pi_i$ be the probability that any changed environment becomes $i$, giving}
\P = (1-\sigma) \piv \evt  + \sigma \I.
}
Since $\piv \evt $ is a rank-one matrix, $\lambda_i(\piv\evt) = 0$ for $i=2, \ldots, n$ \citep[p. 62]{Horn:and:Johnson:1985}.  Hence $\lambda_i(\P) = \sigma$ for $i = 2, \ldots, n$.  

For the case $\tau =1$, then $\lambda_i(\P)= \sigma = 0$,  $i = 2, \ldots, n$, so \eqref{eq:drho-y2} evaluates to 
\ab{
\df{\rho(\M \D) }{m} 
& = \sum_{i=1}^n \lambda_{Pi} (\lambda_{Qi} - 1) y_i^2 \\
& =  1 (1 - 1) y_1^2 + \sum_{i=2}^n 0 (\lambda_{Qi} - 1) y_i^2 = 0.
}
For the case $\tau > 1$, then $\lambda_i(\P) = \sigma > 0$,  $i = 2, \ldots, n$, so \eqref{eq:drho-y2} evaluates to 
\ab{
\df{\rho(\M \D) }{m} 
& = \sum_{i=1}^n \lambda_{Pi} (\lambda_{Qi} - 1) y_i^2 \\
& =  1 (1 - 1) y_1^2 + \sum_{i=2}^n  \sigma (\lambda_{Qi} - 1) y_i^2 < 0,
}
since $\D \neq c \, \I$ for any $c \in \Reals$ implies that $\lambda_{Qi} - 1 < 0$ for $i \in \{2, \ldots, n\}$, and by \ref{item:y not 0}. in the proof of Theorem \ref{Theorem:Main}, $y_i \neq 0$ for some  for $i \in \{2, \ldots, n\}$. \qed
}	

\subsection{Proof of Theorem \ref{Theorem:Cues}, Conditional Dispersal.}\label{section:proof:Cues}
\enumlist{
\item
Basic identities used are $\pfinline{\C}{C_ \kappa} = \D_{\e_\kappa}$, $\evt \e_\kappa = 1$, and $\P \e_\kappa = [\P]_\kappa$.  The derivative formula \eqref{eq:Derivative},
$\partial{\rho(\A)}/\partial{\beta} = \uv(\A)\tr (\partial{\A}/\partial{\beta}) \vv(\A)$ \citep[Sec. 9.1.1]{Caswell:2000}, with respect to parameter $\beta$,
is applied to yield
\an{
&\pf{}{C_\kappa}  \rho(\M\D) = \pf{}{C_\kappa}  \rho(\P [ (\I - \C) + \piv \evt \C] \D) \notag
\notag\\&= \uv\tr \pf{\M}{C_\kappa} \D \vv 
= \uv\tr \P  (-\D_{\ev_\kappa} + \piv \evt \D_{\ev_\kappa})  \D \vv  \notag\\&
= \uv\tr (\P  \piv \evt  \ev_\kappa- \P  \ev_\kappa ) \, D_\kappa \,  v_\kappa\label{eq:Derivkappa} \\&
=  \uv\tr (  \piv -   [\P]_\kappa )  D_\kappa \,  v_\kappa 
=  D_\kappa \,  v_\kappa \sum_{i=1}^n u_i (  \pi_i -   P_{i\kappa} ) \notag\\&
= D_\kappa  \, v_\kappa \, n\, [ \Cov(u_i, \pi_i ) -  \Cov(u_i, P_{i\kappa} )]. \label{Cov-Cov}
}
Note: $1/n^2$ in the covariance terms cancels in \eqref{Cov-Cov}, e.g.
\ab{
\Cov(u_i, P_{i\kappa}) &
= \frac{1}{n} \sum_i u_i P_{i\kappa} - \frac{1}{n^2}\sum_i u_i \sum_i P_{i\kappa} \\&
= \frac{1}{n} \sum_i u_i P_{i\kappa} - \frac{1}{n^2}. 
} 
\item There is always at least one environment in which increased dispersal is advantageous, and at least one  environment where decreased dispersal is advantageous:  If no environment selects for increased dispersal, that means $\partial \rho(\M\D) / \partial{C_\kappa} \leq 0$ for all $\kappa$, hence $\uv\tr (  \piv -   [\P]_\kappa ) \leq 0$, or, combined, $ \uvt (  \piv \evt -  \P) \leq \0\tr$.  Then $\uvt (  \piv \evt -  \P) \piv \leq 0$.  But 
\ab{
 \uvt (  \piv \evt -  \P) \piv = \uvt \piv - \uvt \piv = 0, 
 }
so $\uvt (  \piv -   [\P]_\kappa ) = 0$ for all $\kappa$, which implies either $\piv = [\P]_\kappa\ \forall \ \kappa$ since $\uv > \0$, or $\uvt = \evt$ which requires $\D = c \, \I$ for some $c \in \Reals$.  If neither $\piv = [\P]_\kappa\ \forall \ \kappa$ nor $\D = c \, \I$, then there must be some $\kappa$ for which $\uv\tr (\piv - [\P]_\kappa ) > 0$, hence $\pf{}{C_\kappa} \rho(\P [ (\I - \C) + \piv \evt \C] \D) > 0$.  The parallel argument follows when $\leq$ is replaced by $\geq$ above. 

{\bf Remark}.  When $n=2$, then it must be the case that the spectral radius is maximized at either $(C_1, C_2) =(1,0)$, or at $(C_1, C_2) =(0,1)$.  This is illustrated in the numerical example in Figure \ref{fig:ConditionalDispersal}.

\item A row vector is made from \eqref{eq:Derivkappa}, over $\kappa$:
\ab{
\Gradient_\C \ \rho(\M\D) & \eqdef \matrx{\pf{\rho(\M\D)}{C_\kappa}}_n^{\kappa=1}
=  \matrx{ \uv\tr \pf{\M}{C_\kappa} \D \vv}_n^{\kappa=1} \\&
= \matrx{\uv\tr  ( \P \piv \evt-   \P) \ev_\kappa D_\kappa \  v_\kappa }_n^{\kappa=1}\\&
=  \uv\tr (\piv \vv\tr  -  \P \D_\vv ) \D.
}
\item Since $\Nc \eqdef \{\xiv \suchthat \Gradient_\C \, \rho(\M\D)\  \xiv = 0 \}$ is defined by a single constrain, it is an $n-1$ dimensional linear subspace.  Verification is given that $\zetav = \D_v^{-1} \D^{-1} \piv \in \Nc$:
\ab{
&\Gradient_\C \ \rho(\M\D) \  \zetav = \uv\tr (\piv \vv\tr  -  \P \D_\vv ) \D  (\D_v^{-1} \D^{-1} \piv) \\&
= \uv\tr (\piv \vv\tr  \D  \D_v^{-1} \D^{-1} \piv  -  \P   \piv ) 
= \uv\tr (\piv \vv\tr  \D_v^{-1}  \piv -    \piv ) \\&
= \uv\tr (\piv \ev\tr \piv -   \piv ) 
= \uv\tr (\piv  -   \piv ) = 0.
} \qed
}	

\section{Discussion}

There are two sets of take-home messages from the results here: one, content, and the other, methodology.  Some of the content can be summarized simply as, ``the results of \citet{\McD} generalize to $n$ environments,'' but in that generalization, new relationships emerge that were not visible from the parameters with only 2 environments.  

The content in general may be understood to be part of the larger body of literature on the \emph{Reduction Principle} and the departures from it, which include domains reaching from the evolution of recombination to the evolution of cultural traditionalism.  The environmental change model of \McDA\ produces some specific new results for the reduction phenomena. 

We see first that the ``build up of the genotype on good sites'' can be defined precisely as the \emph{fitness-abundance covariance} --- the covariance between the environment-specific growth rate and the excess abundance above what would emerge without differential growth.  The phase of the census --- whether taken before or after dispersal --- is critical to the properties of the fitness-abundance covariance.

Exploration of the stationary state fitness-abundance covariance and its dependence on census phase is made for general combination of stochastic $\M$ and growth rates $\D$ in Theorem 
\ref {Theorem:FACFisher}, and Corollaries \ref {Corollary:rhoFACderiv}, \ref{Corollary:dvDMdm}, 
\ref{Corollary:log_rho}.    The only constraints are that $\M$ be irreducible and growth rates $D_i$ be positive.   Thus, they could just as well apply to mutation/selection balances as to dispersal/growth balances.  Results for specific classes of $\M$ are found in Theorems \ref {Theorem:PositiveFACDM}, \ref{Theorem:CyclicMD}, and \ref{Theorem:CycleCounterexample}, and Corollary \ref {Corollary:McDPhaseTau}.

Theorem \ref {Theorem:FACFisher} shows that the process of dispersal decreases the \fac\ by the variance in growth rates --- a version of Fisher's Fundamental Theorem.  Corollary \ref{Corollary:rhoFACderiv} shows that the derivatives of post-dispersal $\FAC(\M\D)$ and of $\rho(\M\D)$ always have the same sign with respect to any differentiation of $\M$.  Theorem \ref{Theorem:PositiveFACDM} finds that the pre-dispersal fitness-abundance covariance,  $\FAC(\vv(\D\M))$, is \emph{always} positive when $\M$ is the transition matrix of an ergodic reversible Markov chain with all nonnegative eigenvalues, and growth rates differ between environments.  Reversibility is important here, because a counterexample with $\FAC(\vv(\D\M)) < 0$ is found for periodic chains where one environment has a very small growth rate (Theorem \ref {Theorem:CycleCounterexample}).   It is reasonable to conjecture, given the small region of growth rates in which $\FAC(\vv(\D\M)) < 0$, that $\FAC(\vv(\D\M)) > 0$ for all reversible chains regardless of the signs of their eigenvalues.  Cyclic $\M$, on the other hand, always produce a negative post-dispersal \fac, $\FAC(\M\D)$ (Theorem \ref {Theorem:CyclicMD}).

When the growth rate of an environment is increased, then its stationary proportion of the population increases when the census is just \emph{prior} to dispersal (Corollary \ref{Corollary:dvDMdm}).  When the population is censused just \emph{after} dispersal, the relationship can be reversed, as \citet{\McD} discovered, by extreme patterns of environmental change.  Thus, we have a novel implication, for populations near their stationary distribution, that comparison of the abundance relationships before and after dispersal can provide information about the extremity of the environmental change pattern (Corollary \ref{Corollary:McDPhaseTau}).  

A number of results are obtained for a generalization of the \McDT\ model to $n$ environments \eqref {eq:McDgeneral}.  Corollary \ref {Corollary:McDFACderiv} finds , just as \McDA\ do for two environments, that the post-dispersal fitness-abundance covariance, $\FAC(\M\D)$ (which \McDA\ call the ``multiplier effect''), is positive exactly when the reduction principle operates --- i.e. when the growth rate of the population increases from reduced unconditional dispersal.  It is negative when there are departures from reduction.  This correspondence between a negative fitness-abundance covariance and departures from the reduction phenomenon is, however, specific to the model of \McDA\, and not a general property of departures from reduction for operators of the form $\M\D$.  

The field ecologist would want to know how feasible it is to measure the \fac.  Recall that $\M$ can represent a variety of processes.  When $\M$ represents the dispersal probabilities between patches, then $v_i$ represents the portion of the population in patch $i$, and the quantity $\pi_i$ represents the portion that patch $i$ \emph{would have} in the absence of differential growth rates.  Thus $\pi_i$ is not something that actually exists but is a counterfactual.  It may be feasible, however, to estimate $\piv$ by estimating $\M$ from a measurement of the amount of dispersal between each patch (e.g. through mark and recapture experiments), and computing the Perron vector of the resulting estimated $\M$.  

$\M$ has a different meaning in the model of \McDA, where it represent the Markov chain that the environmental states independently follow in all the patches, and $\pi_i$ is simply the portion of patches that are in environmental state $i$, while $v_i$ is the portion of the population in patches of environmental state $i$.  Each of these is an actual quantity that is potentially measurable.  

The expression $\tau_1^{-1} + \tau_2^{-1}$ from \McDT\ is seen in the general case to be a part of the harmonic mean of the expected durations of states in a Markov chain.  The harmonic mean is shown in Lemma \ref{Lemma:HarmonicMean} to be a simple function of the sum of the eigenvalues of the chain's transition matrix.  Thus the condition on $\tau_1^{-1} + \tau_2^{-1}$ discovered by \McDA\ is really a condition on the eigenvalues of the environment transition matrix.

In Theorem \ref{Theorem:Main}, these three entities --- the reduction phenomenon, the harmonic mean of environment durations, and the eigenvalues of the environment transition matrix --- are tied together in the case of   environmental change processes that are reversible Markov chains.  A sufficient condition for departures from reduction (selection for \emph{increased} unconditional dispersal) is that all of the non-Perron eigenvalues of the environment transition matrix be negative, which represents an extreme pattern of change, in which the harmonic mean of environment durations is less than $1+1/(n-1)$, where $n$ is the number of environments.  This means the environment changes almost every generation.  

This departure from reduction identified by \McDA\ and generalized in Theorem \ref{Theorem:Main} provides a new example summarized by the ``principle of partial control'' \citep{Altenberg:1984}.  The `partial control' in Theorem \ref{Theorem:Main} is that while the organism can control the transformation of its location (i.e.\ dispersal), it cannot control the transformations that change its environment.  

Theorem \ref{Theorem:Main} shows that a sufficient condition for the reduction phenomenon (selection for reduced unconditional dispersal) is that all eigenvalues of the environment transition matrix be positive, corresponding to less extreme environmental change.  A general treatment of the intermediate case --- of mixed positive and negative eigenvalues --- remains an open question.  But Corollary \ref {Corollary:DistanceFromIdentity} shows that there is always some intermediate level of environmental change below which the reduction principle operates.  The reduction principle would be expected to operate for more common patterns of environmental change.

Theorem \ref {Theorem:HouseOfCards} shows that this complexity of behavior disappears when the process of environmental change does not have any causal connection between the identity of sequential environments.  In this case, only the reduction principle operates.

\McDA's model of conditional dispersal is here generalized to arbitrary numbers of environments.  Theorem \ref{Theorem:Cues} shows that conditional dispersal provides another situation where we observe departures from the reduction principle.  Conditional dispersal is mathematically analogous to directed mutation.  Theorem \ref{Theorem:Cues} finds that there is always some environment from which it pays to increase dispersal, provided that there is: 1) some level of environmental change, 2) a causal connection between the current and next environments, and 3) different growth rates among environments.  Therefore, philopatry is not the global evolutionarily stable state.  
This result holds for arbitrary environmental change Markov chains, and holds whether or not \emph{unconditional} dispersal follows the reduction principle.

This seems to contradict the conclusion of \McDT\ that there are ``conditions
under which reliable, cost-free cues to habitat quality, which might
intuitively influence optimal dispersal decisions, should be ignored in
favour of blind natal philopatry.''  This contradiction is resolved by examining the complete adaptive landscape for the conditional dispersal rates (the $n=2$ case in Figure \ref{fig:ConditionalDispersal}).

We see that the evolutionarily stable state of dispersal is highly sensitive to any genetic or phenotypic constraints placed on the range of dispersal combinations.  The hypothesized error rate for environmental cues in the \McDA\ model can constrain the variation to a region where the population growth rate is maximized by philopatry.  But a slight decrease in the error rate can shift the evolutionarily stable state to maximize conditional dispersal from one environment, as shown by \McDA.

Other patterns of phenotypic constraint can be envisioned, and the sensitivity of the ESS in this model to phenotypic constraints leads to a diversity of potential phenomena: intermediate ESS states, bimodal states, or a general condition of evolutionary volatility.  The evolutionary outcome becomes highly dependent on the variational properties \citep{Altenberg:1995:GGEGPM} of the organism.  To the extent that the \McDA\ model of random environments applies to the real world, the results suggest that empirical studies of the evolution of dispersal should find volatile relationships between an organism's dispersal behavior, the variational properties of its dispersal phenotype, and the pattern of environmental change its lineage has experienced.

\subsection{Mathematical Methods}
The second set of take-home messages from this paper regards the mathematical methods.  The primary message is that techniques from the Reduction Principle literature and contemporary linear algebra allow one to obtain analytical results in greater generality than is often pursued.  The common restriction to $2 \times 2$ matrices can be dropped for many results.  

There is the added benefit from generalizing $2 \times 2$ models to the $n \times n$ case, which is that one is forced to see beyond the four particular entries of the $2 \times 2$ matrices to their deeper underlying structures, in particular their eigenvalues and eigenvectors, covariances, and the variational structure of the matrices.  In the case of \citet[online Appendix A, Theorem A]{\McD}, the set of inequalities on the particular vector elements can be unified by a single inequality on a covariance expression, as in Corollary \ref{Corollary:McDFACderiv}.  It is hoped that the tractability of many results for general $n$, and the insights provided from such results, will encourage this approach more widely.

Tractability for Theorems \ref {Theorem:PositiveFACDM}, \ref{Theorem:Main}, \ref {Theorem:HouseOfCards}, and \ref{Theorem:SpectralRadius}, and Corollaries \ref {Corollary:n=2}, \ref{Corollary:SpecialCases}, and \ref{Corollary:DistanceFromIdentity} requires the assumption that the environments form a reversible Markov chain.  The transition matrices of reversible Markov chains are synonymous with symmetrizable stochastic matrices.  The tractability provided by symmetrizable stochastic matrices is the key tool adopted from Karlin's Theorem 5.1 \citeyearpar{Karlin:1982} and \citet[Theorem 4.1]{Friedland:and:Karlin:1975}.  Karlin's Theorem 5.1 appears to have never been used since its publication until it was applied to the analysis of the evolution of mutation rates at multiple loci in \citep{Altenberg:2009:Mutation}.  Recently, however, symmetrizable stochastic matrices have been used by \cite{Schreiber:and:Li:2011:Periodic} to analyze the evolution of dispersal in cyclic environments.

The environmental cycling that produces a departure from the reduction principle in the model of \citet {Schreiber:and:Li:2011:Periodic} satisfies the same condition of extreme environmental change as in Theorem \ref{Theorem:Main}, and the matrices are symmetrizable as well.  But it is fundamentally a different model in that the environments change synchronously throughout all the patches, not independently as in \eqref{eq:McDgeneral}, so it is represented by \eqref {eq:TimeVaryingD}, $\z(t+2) = \M \D_{(2)} \M \D_{(1)} \z(0)$.  Nevertheless, the parallels in its behavior with that of the \McDA\ model are intriguing.  Recall that \citet{Karlin:1982} represented periodic environments by using cyclic matrices, so the model of \citet {Schreiber:and:Li:2011:Periodic} can be represented as 
\begin{eqnarray*}
\lefteqn{\Bmatr{\0 & \M(m)  \\ \M(m)  & \0} 
\Bmatr{\D_{(1)} & \0 \\ \0 & \D_{(2)}}} \\
&= \left( (1-m) \Bmatr{\0 & \I \\ \I  & \0} + m \Bmatr{\0 & \P \\ \P  & \0} \right) 
\Bmatr{\D_{(1)} & \0 \\ \0 & \D_{(2)}}.  &
\end{eqnarray*}
We note that this has the form $(1-m) \A + m \B$ \eqref {eq:PartialControl} from the open problem posed in \citet{Altenberg:2004:Open}, and therefore provides another set of conditions on $\A$ and $\B$ that produce departures from reduction.

The case of general Markov chains remains an open problem for the above results.  The principle difference when considering general Markov chains is that the non-Perron eigenvalues may come in complex-conjugate pairs, which represent cycles of states that are more probable in one direction than the reverse.  Whether directional cycles of the environments can produce any new phenomena for the evolution of dispersal is here an open question.

\subsection{Conclusions}

\citet{Andrewartha:1961} classically defined ecology as ``the scientific
study of the distribution and abundance of organisms.''  In this respect, the \fac\ investigated here is a basic quantity for ecology.  

What makes its behavior more complex than intuition would suggest is that differential growth rates between patches or environments can interact with the multitude of possible dispersal, environmental change, and other mixing processes to produce novel relationships.  The relationships identified by \McDT\ between the \fac, the temporal properties of environmental change, and selection for or against dispersal provided the motivation for the present study.  

The goal here has been to pursue the mathematics underlying these relationships.  In so doing, these relationships are shown to connect to the body of work in the population genetics literature on the Reduction Principle for the evolution of genetic systems and migration, and provide new examples of departure from reduction.  The common mathematics underlying all of these models may lead to the eventual development of a unified theoretical treatment in which the different ecological and evolutionary phenomena are seen as different aspects of a single phenomenological structure.

\section*{Acknowledgements}
I thank Shmuel Friedland for inviting me to speak on work related to Karlin's theorems at the 16th International Linear Algebra Society Conference in Pisa, June, 2010, which was a mathematical feast, and let me meet many of the authors cited here, including Profs. Friedland, Caswell, Horn, Johnson, Li, Kirkland, and Neumann.  I thank Laura Marie Herrmann for assistance with the literature search.  I thank \href{http://www.sciencedaily.com/releases/2011/02/110228104113.htm} {\emph{Science Daily}} \citeyearpar{ScienceDaily:2011-3-1:Stupid} for the coverage of \citet{\McD} that brought their paper to my attention. 

\end{document}